\documentclass[twocolumn]{aastex63}

\usepackage{amsmath}

\shorttitle{Deuterated Molecules in G14.49}
\shortauthors{Sakai et al.}
\begin{document}

\title{The ALMA Survey of 70 $\mu$m Dark High-mass Clumps in Early Stages (ASHES). \\
V. Deuterated Molecules in the 70 $\mu$m dark IRDC G14.492-00.139}

\correspondingauthor{Takeshi Sakai}
\email{takeshi.sakai@uec.ac.jp}

\author[0000-0003-4521-7492]{Takeshi Sakai}
\affiliation{Graduate School of Informatics and Engineering, The University of Electro-Communications, Chofu, Tokyo 182-8585, Japan}

\author[0000-0002-7125-7685]{Patricio Sanhueza}
\affiliation{National Astronomical Observatory of Japan, National Institutes of Natural Sciences, 2-21-1 Osawa, Mitaka, Tokyo 181-8588, Japan}
\affiliation{Department of Astronomical Science, The Graduate University for Advanced Studies, SOKENDAI, 2-21-1 Osawa, Mitaka, Tokyo 181-8588, Japan}

\author[0000-0002-2026-8157]{Kenji Furuya}
\affiliation{National Astronomical Observatory of Japan, National Institutes of Natural Sciences, 2-21-1 Osawa, Mitaka, Tokyo 181-8588, Japan}

\author[0000-0002-8149-8546]{Ken'ichi Tatematsu}
\affiliation{Nobeyama Radio Observatory, National Astronomical Observatory of Japan, National Institutes of Natural Sciences, 462-2 Nobeyama, Minamimaki, Minamisaku, Nagano 384-1305, Japan}
\affiliation{Department of Astronomical Science, The Graduate University for Advanced Studies, SOKENDAI, 2-21-1 Osawa, Mitaka, Tokyo 181-8588, Japan}

\author[0000-0003-1275-5251]{Shanghuo Li}
\affiliation{Korea Astronomy and Space Science Institute, 776 Daedeokdae-ro, Yuseong-gu, Daejeon 34055, Republic of Korea}

\author[0000-0003-3283-6884]{Yuri Aikawa}
\affiliation{Department of Astronomy, School of Science, The University of Tokyo, 7-3-1 Hongo, Bunkyo, Tokyo 113-0033, Japan}

\author[0000-0003-2619-9305]{Xing Lu}
\affiliation{Shanghai Astronomical Observatory, Chinese Academy of Sciences, 80 Nandan Road, Shanghai 200030, People’s Republic of China}
\affiliation{National Astronomical Observatory of Japan, National Institutes of Natural Sciences, 2-21-1 Osawa, Mitaka, Tokyo 181-8588, Japan}

\author[0000-0003-2384-6589]{Qizhou Zhang}
\affiliation{Center for Astrophysics, Harvard \& Smithsonian, 60 Garden Street, Cambridge, MA 02138, USA}

\author[0000-0002-6752-6061]{Kaho Morii}
\affiliation {Department of Astronomy, Graduate School of Science, The University of Tokyo, 7-3-1 Hongo, Bunkyo-ku, Tokyo 113-0033, Japan}
\affiliation {National Astronomical Observatory of Japan, National Institutes of Natural Sciences, 2-21-1 Osawa, Mitaka, Tokyo 181-8588, Japan}

\author[0000-0001-5431-2294]{Fumitaka Nakamura}
\affiliation{National Astronomical Observatory of Japan, National Institutes of Natural Sciences, 2-21-1 Osawa, Mitaka, Tokyo 181-8588, Japan}
\affiliation{Department of Astronomical Science, The Graduate University for Advanced Studies, SOKENDAI, 2-21-1 Osawa, Mitaka, Tokyo 181-8588, Japan}
\affiliation{Department of Astronomy, School of Science, The University of Tokyo, 7-3-1 Hongo, Bunkyo, Tokyo 113-0033, Japan}

\author[0000-0003-2902-2038]{Hideaki Takemura}
\affiliation{Department of Astronomical Science, The Graduate University for Advanced Studies, SOKENDAI, 2-21-1 Osawa, Mitaka, Tokyo 181-8588, Japan}
\affiliation{National Astronomical Observatory of Japan, National Institutes of Natural Sciences, 2-21-1 Osawa, Mitaka, Tokyo 181-8588, Japan}

\author[0000-0003-1604-9127]{Natsuko Izumi}
\affiliation{Academia Sinica Institute of Astronomy and Astrophysics, 11F of Astro-Math Bldg, 1, Section 4, Roosevelt Road, Taipei 10617, Taiwan}

\author[0000-0003-1659-095X]{Tomoya Hirota}
\affiliation{National Astronomical Observatory of Japan, National Institutes of Natural Sciences, 2-21-1 Osawa, Mitaka, Tokyo 181-8588, Japan}
\affiliation{Department of Astronomical Science, The Graduate University for Advanced Studies, SOKENDAI, 2-21-1 Osawa, Mitaka, Tokyo 181-8588, Japan}

\author[0000-0001-9500-604X]{Andrea Silva}
\affiliation{Kavli Institute for the Physics and Mathematics of the Universe, The University of Tokyo, Kashiwa, 277-8583 (Kavli IPMU, WPI) Japan}

\author[0000-0003-0990-8990]{Andr\'{e}s E. Guzm\'{a}n}
\affiliation{National Astronomical Observatory of Japan, National Institutes of Natural Sciences, 2-21-1 Osawa, Mitaka, Tokyo 181-8588, Japan}

\author[0000-0002-3297-4497]{Nami Sakai}
\affiliation{Star and Planet Formation Laboratory, RIKEN Cluster for Pioneering Research, Wako, Saitama 351-0198, Japan}

\author{Satoshi Yamamoto}
\affiliation{Department of Physics, The University of Tokyo, 7-3-1 Hongo, Bunkyo-ku, Tokyo, 113-0033,Japan}

%\author[0000-0003-0990-8990]{Andr\'{e}s E. Guzm\'{a}n}
%\affiliation{National Astronomical Observatory of Japan, National Institutes of Natural Sciences, 2-21-1 Osawa, Mitaka, Tokyo 181-8588, Japan}



\begin{abstract}

We have observed the 70 $\mu$m dark infrared dark cloud (IRDC) G14.492-00.139 in the N$_2$D$^+$ $J$=3--2, DCO$^+$ $J$=3--2, DCN $J$=3--2, and C$^{18}$O $J$=2--1 lines, using the Atacama Large Millimeter/submillimeter Array (ALMA) as part of the ALMA Survey of 70 $\mu$m Dark High-mass Clumps in Early Stages (ASHES). We find that the spatial distribution is different among the observed emission from the deuterated molecular lines. The N$_2$D$^+$ emission traces relatively quiescent regions, while both the DCO$^+$ and DCN emission emanates mainly from regions with signs of active star-formation. In addition, the DCO$^+$/N$_2$D$^+$ ratio is found to be lower in several dense cores than in starless cores embedded in low-mass star-forming regions. By comparing the observational results with chemical model calculations, we discuss the origin of the low DCO$^+$/N$_2$D$^+$ ratio in this IRDC clump. The low DCO$^+$/N$_2$D$^+$ ratio can be explained if the temperature of the dense cores is in the range between the sublimation temperature of N$_2$ ($\sim$20 K) and CO ($\sim$25 K). The results suggest that the dense cores in G14.492-00.139 are warmer and denser than the dense cores in low-mass star-forming regions.

\end{abstract}



\section{Introduction} \label{sec:intro}

High-mass ($>$ 8 $M_\odot$) stars are formed with high accretion rates in massive ($>$10$^3$ $M_\odot$) and dense ($>$ 10$^4$ cm$^{-3}$) clumps. 
However, it is not clear how massive stars are formed in massive clumps. 
A few scenarios have been proposed for high-mass star formation: the turbulent core accretion model \citep{2003ApJ...585..850M}, the competitive accretion model \citep{2001MNRAS.323..785B}, global hierarchical collapse \citep{2019MNRAS.490.3061V}, and the inertial-inflow model \citep{2020ApJ...900...82P}. 
Among these models, initial conditions of high-mass star formation (e.g., initial core mass) are predicted to be different, so that revealing the initial conditions is crucial for understanding the formation process of high-mass stars.

For this purpose, infrared dark clouds (IRDCs) are thought to be good targets.
Although thousands of IRDCs have been found \citep[e.g.,][]{2006ApJ...653.1325S, 2009A&A...505..405P}, 70 $\mu$m dark IRDCs, that are dark in the 70 $\mu$m band due to their low temperatures ($\sim$10 K) \citep[e.g.,][]{2010MNRAS.409...12S, 2015ApJ...815..130G}, are likely to be in the earliest stages of high-mass star formation \citep[e.g.,][]{2013ApJ...773..123S, 2017ApJ...841...97S, 2013ApJ...779...96T, 2018ApJ...861...14C, 2019ApJ...886..130L, 2019A&A...622A..54P}. 
With {\it Herschel} observations, many 70 $\mu$m dark IRDCs, which are likely to be the birthplace of high-mass stars, have been identified \citep{2015ApJ...815..130G, 2015MNRAS.451.3089T}.

Recently, \citet{2019ApJ...886..102S} presented the pilot observations of ASHES (the ALMA Survey of 70 $\mu$m dark High-mass clumps in Early Stages). They conducted observations toward twelve 70 $\mu$m dark IRDCs with ALMA to reveal the initial conditions of high-mass star formation. 
According to their results, there are no massive ($>$ 30 $M_\odot$) prestellar cores in the observed 70 $\mu$m dark IRDCs, while star formation has already started in most of the clumps. They suggested that the dense cores will increase the mass by large-scale collapse, as predicted by clump-fed models such as competitive accretion, global hierarchical collapse, and the inertial-inflow models. 
Furthermore, \citet{Li20b} and \citet{Tafoya21} studied the molecular outflow content of cores from the ASHES survey. In the analysis of the whole pilot survey sample, \citet{Li20b} estimated the accretion rate for the protostars in the 70 $\mu$m dark IRDCs to be less than 3.8$\times$10$^{-6}$ $M_\odot$ yr$^{-1}$.
They suggested that the accretion rate will increase during the star formation process. The ASHES project also includes observations of various molecular lines \citep[see a detailed example on a single ASHES target, IRDC G023.477+0.114 by][]{2021arXiv210901231M}. Among all lines detected in ASHES targets, this paper focuses on the deuterated molecules toward a particular 70 $\mu$m dark IRDC, G14.492-00.139. 

The 70 $\mu$m dark IRDC G14.492-00.139 (hereafter G14.49) is at a distance of 3.9 kpc and has a mass of 5200 $M_\odot$, being the most massive clump in the targets presented in \citet{2019ApJ...886..102S}. These authors identified 37 dense cores and found signs of star formation in 25 of them.  
\citet{2019ApJ...886..130L, Li20b} reported the outflow distribution in G14.49. \citet{Li20b} found six cores driving outflow and that four of them emanate a bipolar outflow.
\citet{Li20b} estimated that the dynamical time of the outflows is 10$^{3-4}$ years.
Thus, G14.49 is globally in an early stage of clump evolution, although star formation has already begun. 

In order to investigate the initial conditions of star formation, observations of deuterated molecules are essential because the abundance of deuterated molecules is thought to increase at the cold starless phase \citep[e.g.,][]{1999ApJ...523L.165C}.
In particular, observations of deuterated ionic molecules are essential for revealing the properties of cold dense gas because the abundance of the deuterated ionic molecules is thought to be sensitive to physical conditions. For example, the abundance of N$_2$D$^+$ would decrease after the temperature raise due to star formation, so that the N$_2$D$^+$ emission can selectively trace the cold starless phase.  On the other hand, the deuterated neutral molecules freeze onto grain surfaces in cold dense gas, and they could be abundant after the onset of star formation due to sublimation from grain surfaces \citep{2012ApJ...747..140S, 2014MNRAS.440..448F}. 
Although the observations of deuterated molecules toward IRDCs have been reported \citep{2010ApJ...713L..50C, 2011A&A...530A.118P, 2013ApJ...773..123S, 2012ApJ...756...60S, 2013ApJ...779...96T, 2016ApJ...821...94K, 2017ApJ...834..193K, 2012ApJ...747..140S, 2015ApJ...803...70S, 2021ApJ...912L...7L}, it is not clear how the distributions and evolutions are different among the deuterated molecules in IRDCs.

In this paper, we report the observational results focused on the DCO$^+$ $J$=3--2, N$_2$D$^+$ $J$=3--2, DCN $J$=3--2, and C$^{18}$O $J$=2--1 lines toward G14.49 with ALMA.
We investigate the differences among the observed deuterated molecules in the 70 $\mu$m dark IRDC. In particular, we focus on the DCO$^+$/N$_2$D$^+$ ratio and investigate how the DCO$^+$/N$_2$D$^+$ ratio varies with physical conditions by comparing the observation results with chemical model calculations. Based on the results, we discuss the properties of dense cores in G14.49.

\section{Observations} \label{sec:obs}

Observations of G14.49 were carried out with ALMA during Cycle 3 (Project ID: 2015.1.01539.S; PI: Sanhueza). 
The 12 m array observations were conducted with 41 antennas, where the baseline ranges from 15 to 330 m. The 7 m array observations and the total power observations were also carried out. We used 8-10 antenna in the 7 m array observations, where the baseline ranges from 8 to 48 m.
The data used in this paper includes the combination of the data of 12 m array, the 7 m array, and the total power. We observed G14.49 with a 10-point mosaic by the 12-m array and a 3-point mosaic by the 7-m array.

We observed the N$_2$D$^+$ $J$=3--2, DCO$^+$ $J$=3--2, DCN $J$=3--2, and C$^{18}$O $J$=2--1 lines in ALMA Band 6 with the dual-polarization mode. The channel width of the spectral windows is 0.17 km s$^{-1}$ for N$_2$D$^+$, DCO$^+$, and DCN, and it is 0.67 km s$^{-1}$ for C$^{18}$O. The observed lines are listed in Table \ref{tbl:line}. 

Data calibrations were carried out by using the Common Astronomy Software Applications (CASA) software package versions 4.5.3, 4.6, and 4.7, while imaging was done using CASA version 5.4 \citep{2007ASPC..376..127M}. To make data cubes, we use an automatic clean script, yclean \citep{2018ApJ...861...14C}. In the cleaning, we used natural weighting.
To avoid artifacts due to the complex structure of the IRDCs emission, we used the multi-scale clean, with scale values of 0, 3, 10, and 30 times of the image pixel size (0.2").
The angular resolution of the images is $\sim$1$^{\prime\prime}$.2, which corresponds to $\sim$0.02 pc at a distance of 3.9 kpc. 
The noise level is listed in Table \ref{tbl:line}.
More details on the data reduction can be seen in \citet{2019ApJ...886..102S} and Contreras et al. (2021, submitted).

\begin{deluxetable}{crrrrr}
\tabletypesize{\scriptsize}
\tablecaption{Observed Lines \label{tbl:line}}
\tablewidth{0pt}
\tablehead{
\colhead{Line} &\colhead{Frequency}  & \colhead{$E_u$/$k$} & \colhead{$\mu^2 S$\tablenotemark{a}} & \colhead{beam size} & \colhead{rms}\\
 & \footnotesize[GHz] & \footnotesize[K]& \footnotesize[D$^2$] &  & \footnotesize[mJy beam$^{-1}$]\\
}
\startdata
N$_2$D$^+$ $J$=3--2 & 231.3218283 & 22.20 & 312.1& 1.4$^{\prime\prime}$$\times$1.0$^{\prime\prime}$ & 11\tablenotemark{b}\\
DCO$^+$ $J$=3--2 & 216.1125822 & 20.74 & 45.6 & 1.5$^{\prime\prime}$$\times$1.1$^{\prime\prime}$ & 9\tablenotemark{b} \\
DCN $J$=3--2 & 217.2385378 & 20.85 & 80.5 & 1.7$^{\prime\prime}$$\times$1.2$^{\prime\prime}$ & 9\tablenotemark{b}   \\
C$^{18}$O $J$=2--1 & 219.5603541 & 15.81 & 0.024 & 1.5$^{\prime\prime}$$\times$1.0$^{\prime\prime}$ & 8\tablenotemark{c}  \\
\enddata
\tablenotetext{a}{from the Cologne Database for Molecular Spectroscopy (CDMS) \citep{2005JMoSt.742..215M}.}
\tablenotetext{b}{rms noise level with the velocity width of 0.17 km s$^{-1}$.}
\tablenotetext{c}{rms noise level with the velocity width of 0.67 km s$^{-1}$.}
\end{deluxetable}

\section{Results} 

\subsection{Integrated Intensity Maps} 

\subsubsection{N$_2$D$^+$, DCO$^+$ and DCN} 
\label{sec:cat}

\begin{figure*}
\epsscale{1.1}
\plotone{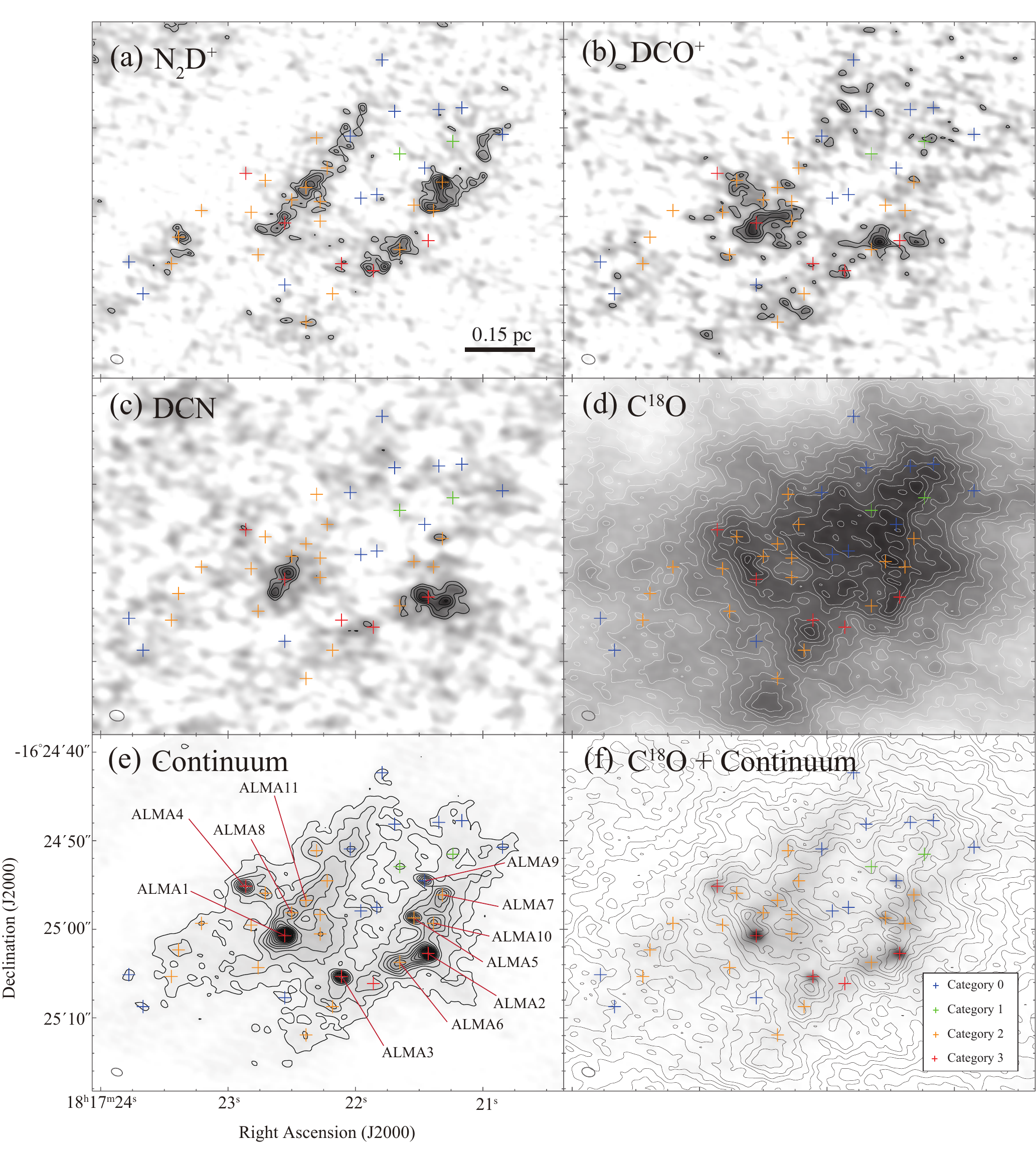}
\caption{Integrated intensity maps of N$_2$D$^+$ $J$=3--2 (a), DCO$^+$ $J$=3--2 (b), DCN $J$=3--2 (c) and C$^{18}$O $J$=2--1 (d) toward G14.492-00.139. The integrated velocity range is from 36 km s$^{-1}$ to 43 km s$^{-1}$. Contour levels start from 3$\sigma$ and increase in steps of 1$\sigma$ for (a), (b), and (c) [(a) $1\sigma=18$ mJy beam$^{-1}$ km s$^{-1}$, (b) $1\sigma=15$ mJy beam$^{-1}$ km s$^{-1}$, (c) $1\sigma=17$ mJy beam$^{-1}$ km s$^{-1}$]. For (d), contour levels start from 3$\sigma$ and increase in steps of 3$\sigma$ [$1\sigma=16$ mJy beam$^{-1}$ km s$^{-1}$]. (e) Continuum image. (f) C$^{18}$O integrated intensity map (contours) overlaid on the continuum image (grey scale). The eleven cores with high H$_2$ column density are labeled as ALMA1-ALMA11. Cross marks indicate the positions of the dense cores identified by \citet{2019ApJ...886..102S}. The four categories of the cores are shown in different colors [red: category 3, orange: category 2, green: category 1, blue: category 0], see Section~\ref{sec:cat} for the category definition.
\label{fig:imaps}}
\end{figure*}

\begin{figure*}
\epsscale{1.0}
\plotone{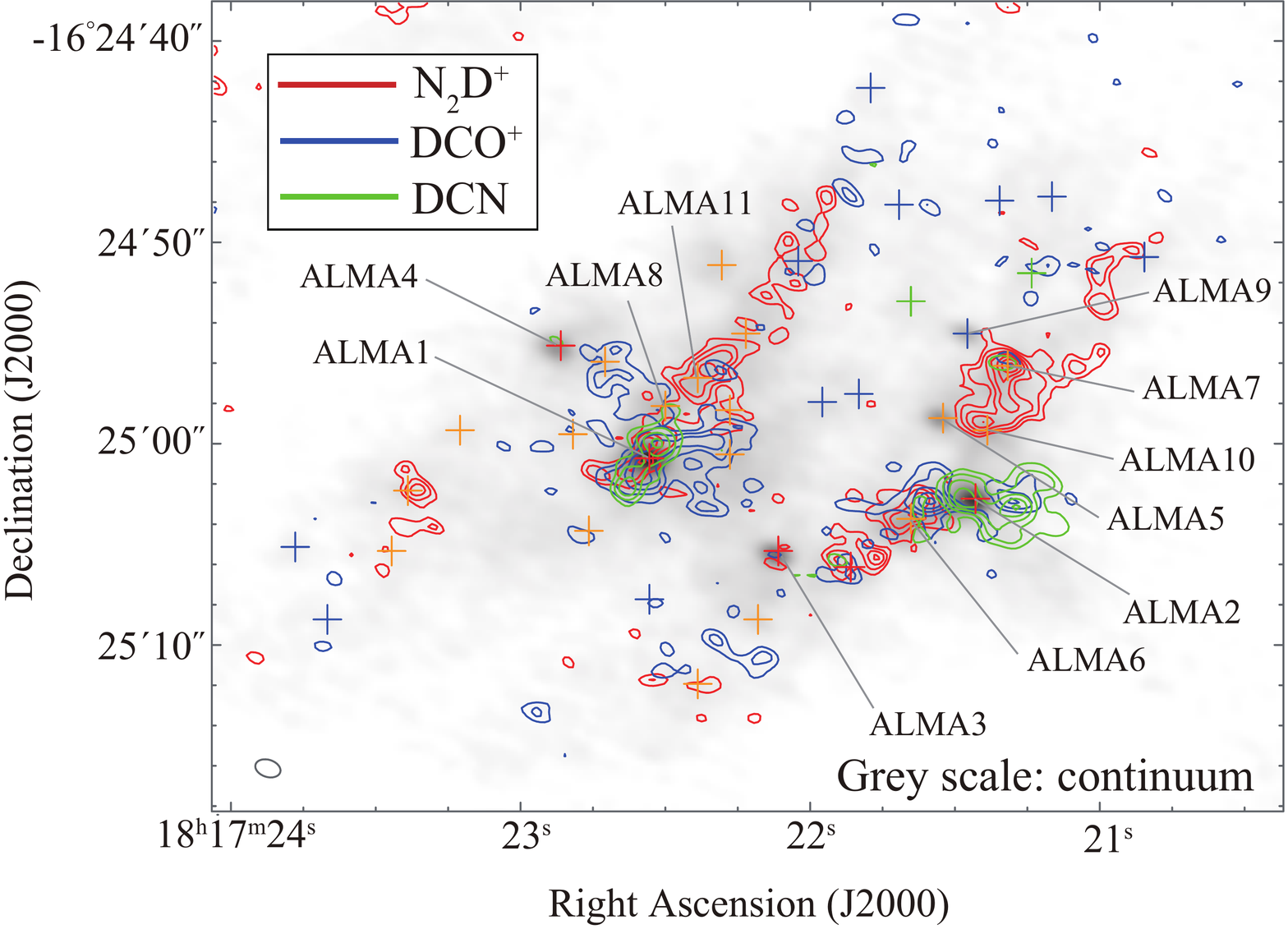}
\caption{Integrated intensity maps of N$_2$D$^+$ $J$=3--2 (red contours), DCO$^+$ $J$=3--2 (blue contours) and DCN $J$=3--2 (green contours) superposed on the continuum image (grey). 
Cross marks indicate the positions of the dense cores identified by Sanhueza et al. (2019). See the caption in Figure 1.\label{fig:imapD}}
\end{figure*}

Figures \ref{fig:imaps}a-c show integrated intensity maps of the N$_2$D$^+$ $J$=3--2, DCO$^+$ $J$=3--2 and DCN $J$=3--2 lines toward G14.49. 
The velocity range of the integration is from 36 km s$^{-1}$ to 43 km s$^{-1}$.
In Figures \ref{fig:imaps}a-c, the spatial distribution is different among the molecules. The N$_2$D$^+$ emission is seen along the two parallel filamentary structures stretching along the southeast-northwest direction, while the DCO$^+$ and DCN emission comes mainly from the southern region of this clump (see Figure \ref{fig:imapD}).

In Figure \ref{fig:imaps}, we plot the positions of the dense cores identified by \citet{2019ApJ...886..102S} as cross marks. \citet{2019ApJ...886..102S} classified dense cores into four categories; 0: prestellar cores (blue), 1: star-forming cores detected only with molecular outflow emission (green), 2: star-forming cores detected only with warm-core line emission (orange), 3: star-forming cores detected with molecular outflow and warm-core line emission (CH$_3$OH and H$_2$CO) (red). Among the four categories, the cores in category 3 are thought to be the most evolved. 
In Figure \ref{fig:imaps}e, the eleven cores with high H$_2$ column density are labeled as ALMA1-ALMA11 (ALMA1 is the most massive core). The four highest column density cores are classified as category 3, and only one core, ALMA9, is classified as category 0 (starless core).
The properties of the cores are summarized in Table \ref{tbl:CoreP}.

In Figure \ref{fig:imaps}, it appears that all the cores in category 3 are located in the southern region of this clump. In contrast, many of the prestellar cores (category 0) are located in the northern region.
The DCN emission is distributed near the cores in category 3. The integrated intensity of the DCO$^+$ emission is relatively strong near the cores in category 3. On the other hand, the N$_2$D$^+$ emission is extended in the northern quiescent regions (Figure \ref{fig:imapD}).
Thus, star formation seems to affect the spatial distribution of the deuterated molecules.

\subsubsection{C$^{18}$O} 

Figure \ref{fig:imaps}d shows the integrated intensity map of the C$^{18}$O $J$=2--1 line. 
The C$^{18}$O emission is found to be spread throughout the whole clump. 
Although the morphology of the C$^{18}$O emission is similar to that of the continuum emission, there is no C$^{18}$O peak toward most of the continuum peaks (Figure \ref{fig:imaps}f).
In addition, the C$^{18}$O emission is found to be relatively strong toward the quiescent northern region of this clump, but uncorrelated with the emission from the deuterated molecules.  
Thus, the C$^{18}$O emission does not seem to trace well the dense regions of G14.49.  

\subsection{Channel Maps} 

\subsubsection{N$_2$D$^+$, DCO$^+$ and DCN} \label{sec:chmapd}

\begin{figure*}
\epsscale{1.1}
\plotone{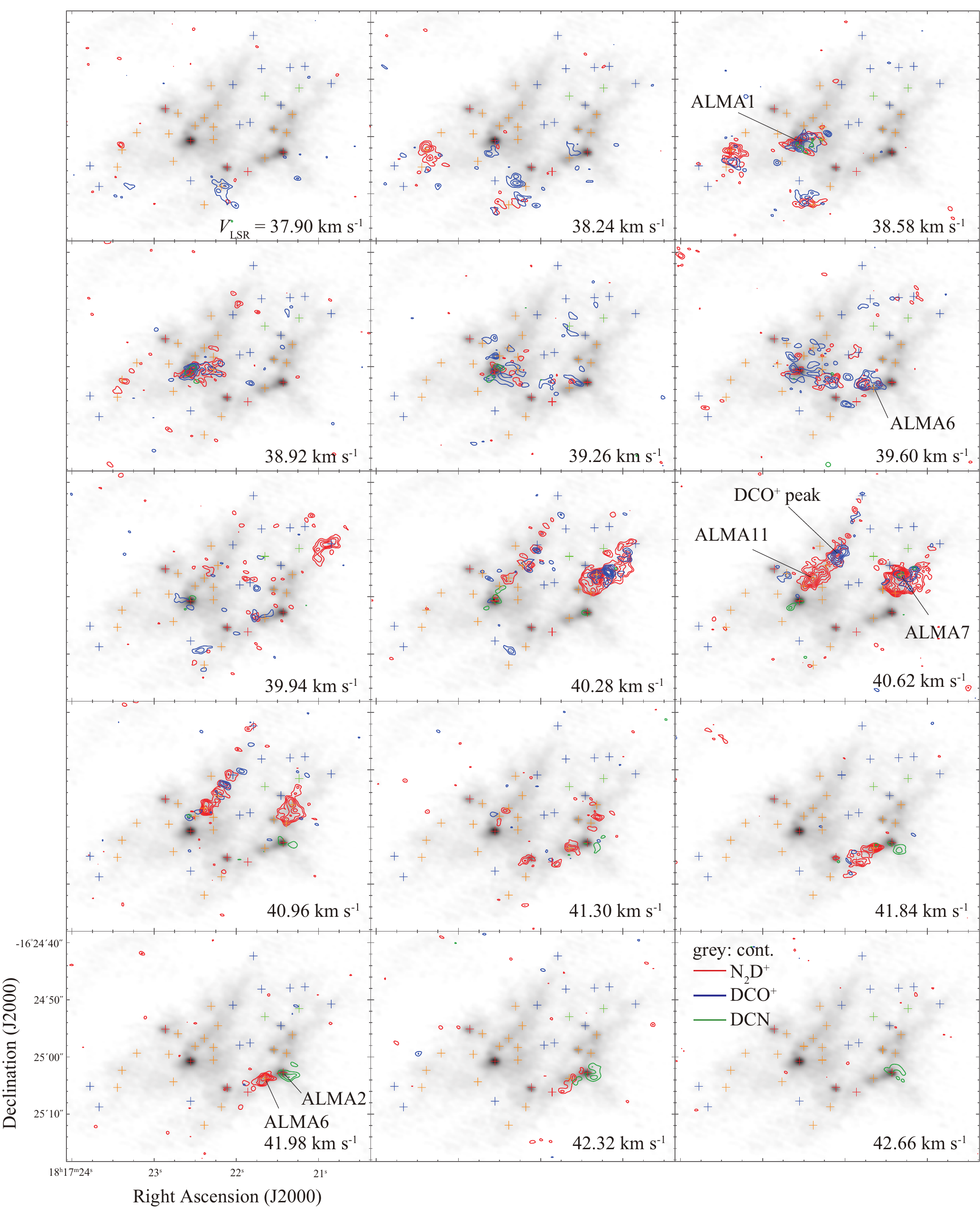}
\caption{Channel maps of N$_2$D$^+$ $J$=3--2 (red contours), DCO$^+$ $J$=3--2 (blue contours), DCN $J$=3--2 (green contours). Contour levels start from 3$\sigma$ and increase in steps of 1$\sigma$ [N$_2$D$^+$: $1\sigma= $ 11 mJy beam$^{-1}$, DCO$^+$: $1\sigma= $ 9 mJy beam$^{-1}$, DCN: $1\sigma= $ 9 mJy beam$^{-1}$]. In each panel, the continuum image is shown in grey scale. Cross marks indicate the positions of the dense cores identified by \citet{2019ApJ...886..102S}. See the caption in Figure 1.\label{fig:chmap}
}
\end{figure*}

Figure \ref{fig:chmap} shows the velocity channel map of the N$_2$D$^+$ $J$=3--2, DCO$^+$ $J$=3--2, and DCN $J$=3--2 lines overlaid on the continuum image as background. Although there are hyperfines in the N$_2$D$^+$ $J$=3--2 line, the broadening of the linewidth due to the hyperfines is only about 0.1 km s$^{-1}$. Thus, the hyperfine splitting has no significant effect in the distributions of the N$_2$D$^+$ emission observed in the channel maps.
In the lower velocity range (37.90-39.60 km s$^{-1}$), the N$_2$D$^+$ and DCO$^+$ emission is detected toward the southern region. The emission from ALMA1, an active star-forming core (category 3), is seen in a relatively wide velocity range.
At channels corresponding to velocities of 39.26 km s$^{-1}$ and 39.60 km s$^{-1}$, the DCO$^+$ emission is more extended than the N$_2$D$^+$ emission.

In the middle-velocity range (39.94-40.96 km s$^{-1}$), we can see relatively strong N$_2$D$^+$ emission toward several dense cores.
In the channels at velocities of 40.28 km s$^{-1}$ and 40.62 km s$^{-1}$, the N$_2$D$^+$ emission is strong toward ALMA7, where the DCO$^+$ emission peak is associated. 
At the velocity of 40.62 km s$^{-1}$, the N$_2$D$^+$ emission is strong toward ALMA11, while the DCO$^+$ and DCN emission is undetected. 

The filamentary structure, which is seen in the N$_2$D$^+$ integrated intensity map (Figure \ref{fig:imaps}a), appears near the center of the maps in the velocity range of 40.28-40.96 km s$^{-1}$, and it has no clear velocity gradient. 
Toward the northern region of this filament, we can see a DCO$^+$ peak without a corresponding dust continuum core.

In the higher velocity range (41.30-43.00 km s$^{-1}$), the N$_2$D$^+$ emission peaks toward ALMA6, where there are no DCO$^+$ and DCN peaks. 
On the other hand, only the DCN emission is detected toward ALMA2 in the velocity range of 40.28-42.66 km s$^{-1}$. The DCN emission is extended toward the east of ALMA2. 
Since an outflow from ALMA2 is reported by \citet{Li20b} (see their Figure 1.2), the extended DCN emission is most likely related to the molecular outflow.

\subsubsection{C$^{18}$O} \label{sec:chmapC18O}

\begin{figure*}
\epsscale{1.1}
\plotone{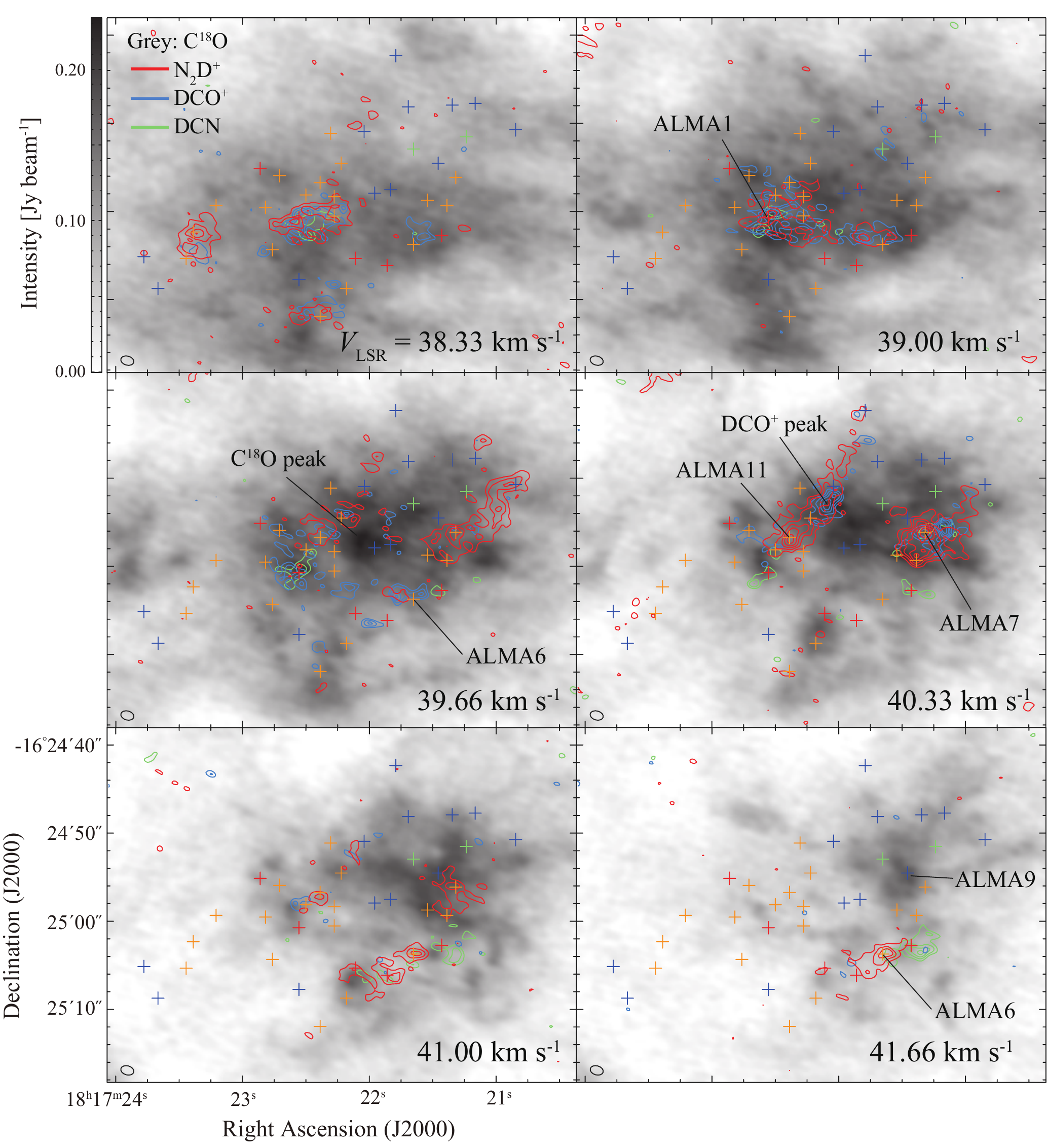}
\caption{Channel maps of C$^{18}$O $J$=2--1 (grey scale), N$_2$D$^+$ $J$=3--2 (red contours), DCO$^+$ $J$=3--2 (cyan contours), DCN $J$=3--2 (green contours). Contour levels start from 3$\sigma$ and increase in steps of 2$\sigma$ for N$_2$D$^+$ and $1\sigma$ for DCO$^+$ and DCN [N$_2$D$^+$: $1\sigma= $ 7 mJy beam$^{-1}$, DCO$^+$: $1\sigma= $ 6 mJy beam$^{-1}$, DCN: $1\sigma= $ 7 mJy beam$^{-1}$]. Cross marks indicate the positions of the dense cores identified by \citet{2019ApJ...886..102S}. See the caption in Figure 1.\label{fig:chmapC18O}
}
\end{figure*}

Figure \ref{fig:chmapC18O} shows channel maps including C$^{18}$O (grey scale), N$_2$D$^+$ (red contours), DCO$^+$ (cyan contours), and DCN (green contours). 
Since the velocity resolution of the C$^{18}$O data is lower than that of the other molecular line data, the N$_2$D$^+$, DCO$^+$, and DCN data are smoothed to the same velocity resolution as that of the C$^{18}$O.

At a velocity of 39.66 km s$^{-1}$, the C$^{18}$O emission is bright toward the center of the clump, although there is no dense core identified with the dust continuum toward the C$^{18}$O peak.
We also find no C$^{18}$O peak toward the N$_2$D$^+$ peaks.
In particular, the C$^{18}$O emission is found to be depleted toward ALMA11 at a velocity of 40.33 km s$^{-1}$.
This could be due to the CO depletion toward the dense core, indicating that the temperature of ALMA11 is lower than the sublimation temperature of CO ($\sim$25 K).

At a velocity of 40.33 km s$^{-1}$, we can see a DCO$^+$ peak located on the northern region of the N$_2$D$^+$ filament. The C$^{18}$O emission is found to be relatively strong near the DCO$^+$ peak, suggesting that CO depletion may not be efficient there.

At a velocity of 41.66 km s$^{-1}$, there is a C$^{18}$O peak toward ALMA9, which is a starless core. The non-detection of the deuterated molecular lines toward ALMA9 could be partly due to a low volume density of this core. The critical density of C$^{18}$O $J$=2--1 ($\sim$10$^4$ cm$^{-3}$) is much lower than those of the observed deuterated molecular lines observed in this study ($\sim$10$^6$ cm$^{-3}$). 
The density estimated from the dust continuum emission is as high as 1.4$\times$10$^7$ cm$^{-3}$ (Table \ref{tbl:CoreP}), where the line-of-sight depth is assumed to be equal to the apparent width of the core. However, this picture is too simplified, because the core might be elongated along the line of sight. Alternatively, it may be possible that the core is chemically very young and deuterated molecules are not yet abundant enough to be detected. In order to investigate this, observations of the normal (not-deuterated) species, such as H$^{13}$CO$^+$ and N$_2$H$^+$, are crucial.

\subsection{Spectra} 

\begin{figure*}
\epsscale{1.1}
\plotone{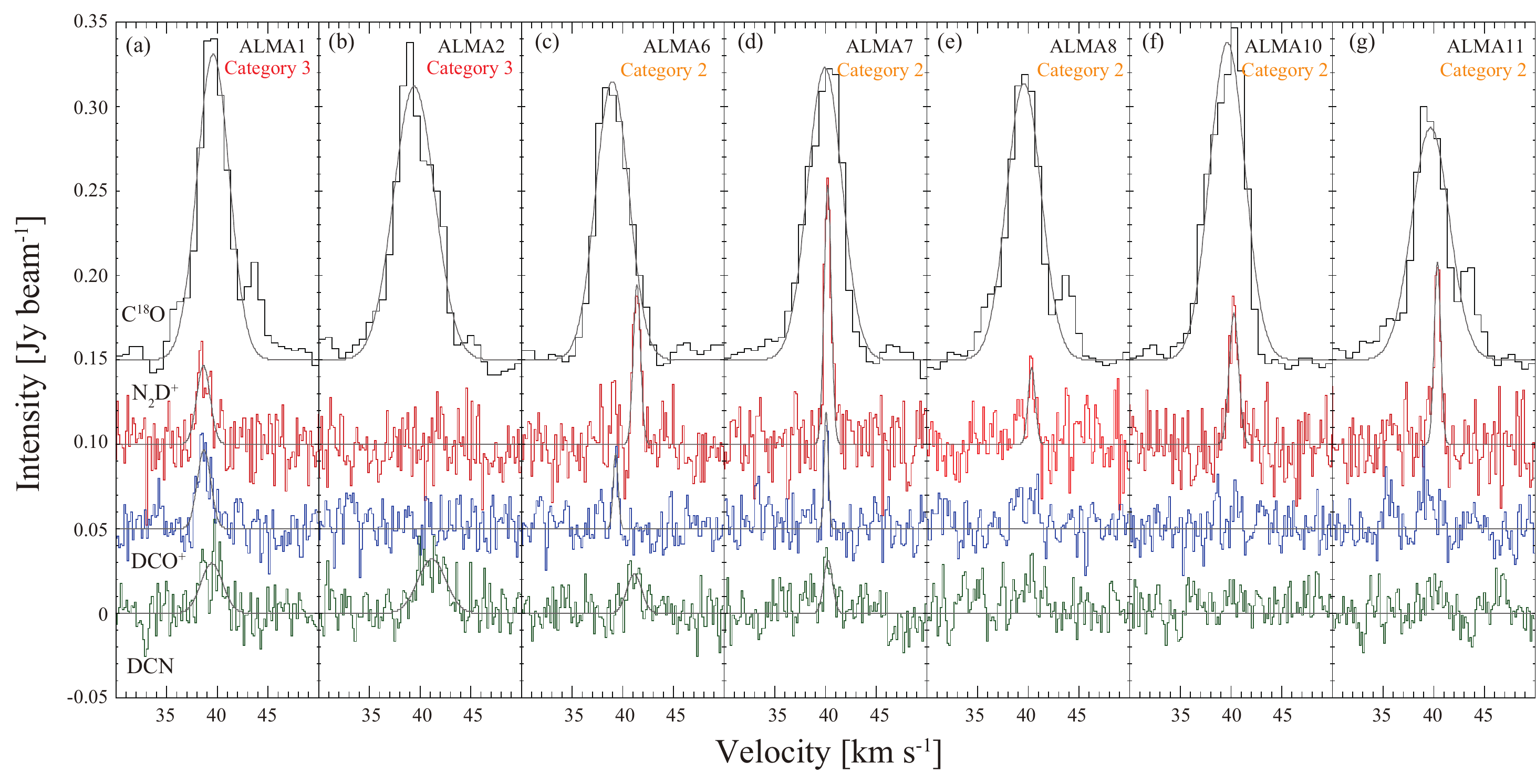}
\caption{Spectra of C$^{18}$O $J$=2--1 (black), N$_2$D$^+$ $J$=3--2 (red), DCO$^+$ $J$=3--2 (blue), DCN $J$=3--2 (green) toward the seven selected cores. The results of single Gaussian fitting are overlaid on the spectra as grey lines.\label{fig:spec}}
\end{figure*}

Figure \ref{fig:spec} shows the spectra of C$^{18}$O, N$_2$D$^+$, DCO$^+$, and DCN toward the peak position of seven selected out of the eleven cores (ALMA1-11) listed in Table \ref{tbl:CoreP}. A criterion of the selection is that N$_2$D$^+$, DCO$^+$, or DCN is detected above the 3 sigma noise level.  The line parameters derived by a single Gaussian fitting are listed in Tables \ref{tbl:N2Dp}-\ref{tbl:C18O}.

\begin{deluxetable}{crrr}
\tabletypesize{\scriptsize}
\tablecaption{Line Parameters of N$_2$D$^+$\label{tbl:N2Dp}}
\tablewidth{0pt}
\tablehead{
\colhead{Core ID} &\colhead{$S_{pk}$}  & \colhead{$\Delta V$} & \colhead{$V_{\rm LSR}$}\\
& \footnotesize[Jy beam$^{-1}$] & \footnotesize[km s$^{-1}$]& \footnotesize[km s$^{-1}$]\\
}
\startdata
ALMA1 & 0.046${\pm 0.007}$ &  1.57${\pm 0.26}$ & 38.63${\pm 0.11}$ \\
ALMA2 & --- & --- & --- \\
ALMA6 & 0.094${\pm 0.009}$ &  0.85${\pm 0.10}$ & 41.44${\pm 0.04}$  \\
ALMA7 & 0.153${\pm 0.009}$ &  0.83${\pm 0.05}$ & 40.22${\pm 0.02}$  \\
ALMA8 & 0.045${\pm 0.009}$ &  0.86${\pm 0.20}$ & 40.13${\pm 0.09}$ \\
ALMA10 & 0.077${\pm 0.008}$ &  1.10${\pm 0.13}$ & 40.29${\pm 0.06}$  \\
ALMA11 & 0.107${\pm 0.010}$ &  0.76${\pm 0.08}$ & 40.36${\pm 0.04}$ \\
\enddata
\tablenotetext{}{$S_{pk}$: peak flux density; $\Delta V$: full width per half maximum}
\end{deluxetable}

\begin{deluxetable}{crrr}
\tabletypesize{\scriptsize}
\tablecaption{Line Parameters of DCO$^+$\label{tbl:DCOp}}
\tablewidth{0pt}
\tablehead{
\colhead{Core ID} &\colhead{$S_{pk}$}  & \colhead{$\Delta V$} & \colhead{$V_{\rm LSR}$}\\
 & \footnotesize[Jy beam$^{-1}$] & \footnotesize[km s$^{-1}$]& \footnotesize[km s$^{-1}$]\\
}
\startdata
ALMA1 & 0.046${\pm 0.004}$ &  1.59${\pm 0.17}$ & 38.69${\pm 0.07}$ \\
ALMA2 & --- & --- & --- \\
ALMA6 & 0.044${\pm 0.008}$ &  0.58${\pm 0.12}$ & 39.30${\pm 0.05}$  \\
ALMA7 & 0.069${\pm 0.009}$ &  0.47${\pm 0.07}$ & 40.07${\pm 0.03}$  \\
ALMA8 & --- & --- & ---  \\
ALMA10 & --- & --- & ---   \\
ALMA11 & --- & --- & ---  \\
\enddata
\tablenotetext{}{$S_{pk}$: peak flux density; $\Delta V$: full width per half maximum}
\end{deluxetable}

\begin{deluxetable}{crrr}
\tabletypesize{\scriptsize}
\tablecaption{Line Parameters of DCN\label{tbl:DCN}}
\tablewidth{0pt}
\tablehead{
\colhead{Core ID} &\colhead{$S_{pk}$}  & \colhead{$\Delta V$} & \colhead{$V_{\rm LSR}$}\\
 & \footnotesize[Jy beam$^{-1}$] & \footnotesize[km s$^{-1}$]& \footnotesize[km s$^{-1}$]\\
}
\startdata
ALMA1 & 0.030${\pm 0.004}$ &  2.54${\pm 0.36}$ & 38.46${\pm 0.15}$ \\
ALMA2 & 0.032${\pm 0.003}$ & 3.29${\pm 0.38}$ & 41.17${\pm 0.16}$ \\
ALMA6 & 0.023${\pm 0.004}$ &  1.73${\pm 0.37}$ & 41.13${\pm 0.16}$  \\
ALMA7 & 0.031${\pm 0.006}$ &  1.04${\pm 0.23}$ & 40.25${\pm 0.10}$  \\
ALMA8 & --- & --- & ---  \\
ALMA10 & --- & --- & ---   \\
ALMA11 & --- & --- & ---  \\
\enddata
\tablenotetext{}{$S_{pk}$: peak flux density; $\Delta V$: full width per half maximum}
\end{deluxetable}

\begin{deluxetable}{crrr}
\tabletypesize{\scriptsize}
\tablecaption{Line Parameters of C$^{18}$O\label{tbl:C18O}}
\tablewidth{0pt}
\tablehead{
\colhead{Core ID} &\colhead{$S_{pk}$}  & \colhead{$\Delta V$} & \colhead{$V_{\rm LSR}$}\\
 & \footnotesize[Jy beam$^{-1}$] & \footnotesize[km s$^{-1}$]& \footnotesize[km s$^{-1}$]\\
}
\startdata
ALMA1 & 0.181${\pm 0.004}$ &  3.74${\pm 0.10}$ & 39.61${\pm 0.04}$ \\
ALMA2 & 0.162${\pm 0.003}$ &  4.76${\pm 0.11}$ & 39.42${\pm 0.05}$ \\
ALMA6 & 0.165${\pm 0.003}$ &  3.86${\pm 0.08}$ & 39.97${\pm 0.04}$  \\
ALMA7 & 0.173${\pm 0.004}$ &  4.00${\pm 0.08}$ & 39.94${\pm 0.04}$  \\
ALMA8 & 0.164${\pm 0.004}$ &  4.01${\pm 0.11}$ & 39.58${\pm 0.05}$ \\
ALMA10 & 0.188${\pm 0.003}$ &  4.09${\pm 0.08}$ & 39.64${\pm 0.04}$ \\
ALMA11 & 0.140${\pm 0.004}$ &  4.38${\pm 0.13}$ & 39.69${\pm 0.05}$ \\
\enddata
\tablenotetext{}{$S_{pk}$: peak flux density; $\Delta V$: full width per half maximum}
\end{deluxetable}

A striking feature in Figure \ref{fig:spec} is that the C$^{18}$O line is broader than the other emission lines.
The full-width at half maximum (FWHM) velocity width of the C$^{18}$O line is typically 4 km s$^{-1}$ (Table \ref{tbl:C18O}).
Since the C$^{18}$O lines show double peaks toward ALMA1, ALMA8, and ALMA11, the broad velocity width could be partly due to the fact that the C$^{18}$O emission traces several cloud components in the line of sight.
The velocity widths of the N$_2$D$^+$ and DCO$^+$ lines are 0.5-1 km s$^{-1}$ toward the cores in category 2 (ALMA6, ALMA7, ALMA8, ALMA10, and ALMA11).
In particular, the FWHM velocity width of DCO$^+$ line is as narrow as 0.5 km s$^{-1}$ for ALMA6 and ALMA7, which is comparable to those of low-mass cores in low-mass star-forming regions (Redaelli et al. 2019).
Although the velocity width of the N$_2$D$^+$ line is slightly broader than that of DCO$^+$, this could be partly due to the hyperfine components of N$_2$D$^+$ $J$=3--2, as mentioned in Section \ref{sec:chmapd}.

Since the critical density of the C$^{18}$O emission is lower than that of the other observed lines and CO depletion is heavier at high-density and low-temperature regions, the C$^{18}$O emission likely traces lower density regions of the clump.
The low-density envelope could be more turbulent than the dense cores. If so, the low velocity width of the deuterated molecular lines is due to the dissipation of turbulence in the dense regions, as suggested by \citet{Li20a, 2021ApJ...912L...7L} toward NGC6334S.

Figure \ref{fig:spec} also shows that the N$_2$D$^+$ emission is stronger than the DCO$^+$ emission toward the cores in category 2.
In particular, the DCO$^+$ emission is not detected above 3 sigma noise level toward ALMA8, ALMA10, and ALMA11.
Toward ALMA6, the N$_2$D$^+$ and DCO$^+$ emission is detected, but their peak velocities are different from each other.
Thus, the core traced by the N$_2$D$^+$ emission toward ALMA6 is likely to be in a similar condition to ALMA8, ALMA10, and ALMA11.
Although there is no C$^{18}$O peak at the velocity of the N$_2$D$^+$ peak toward ALMA6, this could be due to CO depletion onto grain surfaces in the dense core traced by the N$_2$D$^+$ emission.

In this 70 $\mu$m-dark IRDC, the peak intensity of the N$_2$D$^+$ $J$=3--2 emission is higher than that of the DCO$^+$ $J$=3--2 emission, which is different from that observed in cores found in low-mass star-forming regions \citep{2000A&A...356.1039T, 2019A&A...629A..15R}. Thus, the cores in G14.49 may have a different chemical composition from those of the low-mass cores, probably caused by the different environments in high-mass star-forming regions in comparison with low-mass star-forming regions, as will be discussed later in this work.

\section{Discussion} 

\subsection{Different Distributions of Deuterated Species} 

We have found that the spatial distribution of the emission is different among the observed deuterated molecules. Although both the N$_2$D$^+$ and DCO$^+$ molecules are ionic species, we find that the N$_2$D$^+$ emission traces quiescent regions, while the DCO$^+$ emission tends to be relatively strong toward the active star-forming regions.
This difference could be due to the different formation and destruction processes between them.

%The N$_2$D$^+$ molecule is formed by the reactions between N$_2$ and H$_2$D$^+$, and is destroyed by CO following N$_2$D$^+$ + CO $\rightarrow$ N$_2$ + DCO$^+$.

N$_2$D$^+$ is formed by the reactions between N$_2$ and H$_2$D$^+$, and is destroyed by CO or electrons:
\begin{align}
&{\rm N_2} + {\rm H_2D^+} \rightarrow {\rm N_2D^+} + {\rm H_2},  \label{react:n2d+_p} \\
&{\rm N_2D^+} + {\rm CO} \rightarrow {\rm N_2} + {\rm DCO^+}, \label{react:n2d+_d1} \\
&{\rm N_2D^+} + {\rm e^-} \rightarrow {\rm ND} + {\rm N}\,\,{\rm or}\,\,{\rm N_2} + {\rm D}. \label{react:n2d+_d2}
\end{align}

Since CO is depleted onto grain surfaces in cold ($<$ 25 K) regions, the abundance of N$_2$D$^+$ can be enhanced in cold regions, while it decreases in warm regions due to the reaction with CO.

%The DCO$^+$ molecule is formed by CO + H$_2$D$^+$ and is destroyed by an electron or hydrogen atom.

DCO$^+$ is formed by CO + H$_2$D$^+$ or by Reaction \ref{react:n2d+_d1}, 
and is destroyed by electrons:
\begin{align}
&{\rm CO} + {\rm H_2D^+} \rightarrow {\rm DCO^+} + {\rm H_2},  \label{react:dco+_p} \\
&{\rm DCO^+} + {\rm e^-} \rightarrow {\rm CO} + {\rm D}. \label{react:dco+_d}
\end{align}

Thus, the DCO$^+$ abundance does not decrease rapidly even after the sublimation of CO, and it depends weakly on the temperature compared with the N$_2$D$^+$ abundance.
Consequently, the DCO$^+$ emission can also trace relatively warm regions.
This is consistent with the observation results that the DCO$^+$ emission is distributed near the active star-forming regions.
Busquet et al. (2018) reported similar trends of N$_2$H$^+$ and HCO$^+$ in the high-mass star-forming region AFGL 5142.

In addition, DCO$^+$ (and also HCO$^+$) can be destroyed by the molecules with high proton affinity, such as H$_2$O. Thus, the DCO$^+$ abundance can decrease in the hot regions ($T$$>$100 K) near protostars, where H$_2$O and complex organic molecules are liberated from the grain mantles. In fact, there is no DCO$^+$ emission toward ALMA2 (category 3).
However, it cannot be explained only by this picture that the N$_2$D$^+$ emission is stronger than the DCO$^+$ emission toward several dense cores, where the temperature is well below the sublimation temperature of H$_2$O. 
This will be discussed in the next section.

The DCN molecule is likely to be depleted onto grain surfaces in cold and dense regions because the sublimation temperature of DCN (HCN) is about 80 K \citep{1983A&A...122..171Y}.
Thus, the detection of the DCN emission toward dense cores could suggest the existence of hot regions.
In such hot regions, DCO$^+$ and N$_2$D$^+$ are thought to be destroyed by CO, electron, or sublimated molecules with higher proton affinity.
In fact, only the DCN emission is detected toward ALMA2, an active star-forming core (category 3), as mentioned in Section~\ref{sec:chmapd}.
The DCN emission is also detected toward ALMA7 (category 2).  If the above picture of DCN (HCN) is the case, this result implies that a protostar has been born in ALMA7.
In contrast, the DCN emission is not detected toward ALMA11, suggesting no hot regions in ALMA11. Although ALMA11 is classified as category 2, it is likely that active star formation has not yet begun there.

\subsection{The DCO$^+$/N$_2$D$^+$ Ratio}

\subsection{Derivation of DCO$^+$/N$_2$D$^+$} 

We have found that the peak intensity of the N$_2$D$^+$ $J$=3--2 emission is higher than that of the DCO$^+$ $J$=3--2 emission toward several dense cores. As mentioned in Section 3.3, this is opposite to what is observed in low-mass star-forming regions. Since the peak intensity depends on the kinetic temperature, we compare the abundance ratios between the observed cores in G14.49 and the other sources. For this purpose, we derive the DCO$^+$/N$_2$D$^+$ abundance ratio toward the peak of the six cores in which N$_2$D$^+$ and/or DCO$^+$ emission is detected.
As for ALMA6, there are two velocity components (see Figure \ref{fig:chmap} and Figure \ref{fig:spec}), so that we derive the DCO$^+$/N$_2$D$^+$ ratio for each velocity component; the lower velocity component is labeled as ALMA6a, and the higher velocity component is labeled as ALMA6b.

In the derivation of the column densities, we assume local thermodynamic equilibrium (LTE) conditions and optically thin gas. The equations used for the derivation are described in Appendix B.
For the derivation of the DCO$^+$/N$_2$D$^+$ abundance ratio, the same excitation temperature is assumed for DCO$^+$ and N$_2$D$^+$.
Table \ref{tbl:DCON2D} presents the DCO$^+$/N$_2$D$^+$ ratio for the excitation temperature of 20 K, where the errors denoted in Table \ref{tbl:DCON2D} include the observational error (rms noise) and the error of the excitation temperature (10-30 K). The error due to the excitation temperature is less than 7 \%.
Even if the different excitation temperatures were assumed for DCO$^+$ and N$_2$D$^+$ within a temperature range of 10-30 K, the DCO$^+$/N$_2$D$^+$ ratio changes less than 35 \%.

\begin{deluxetable}{crrr}
\tabletypesize{\scriptsize}
\tablecaption{DCO$^+$/N$_2$D$^+$ Abundance Ratio\label{tbl:DCON2D}}
\tablewidth{0pt}
\tablehead{
\colhead{Core ID} &\colhead{[DCO$^+$]/[N$_2$D$^+$]} & \colhead{$N$(N$_2$D$^+$)$_{20 {\rm K}}$} &\colhead{Reference}\\
 &  & \footnotesize[10$^{11}$ cm$^{-2}$] & \\
}
\startdata
ALMA1 &  0.83$_{-0.34}^{+0.34}$ & 8.6 &1 \\
ALMA6a &  $>$1.74 & $<$0.90 & 1\\
ALMA6b &  $<$0.12 & 9.5 & 1\\
ALMA7 &  0.21$_{-0.08}^{+0.08}$ & 15 & 1\\
ALMA8 &  $<$0.34 & 4.6 & 1\\
ALMA10 & $<$0.11 & 10 & 1\\
ALMA11 &  $<$0.13 & 9.7 & 1\\
Total Power & 0.97$_{-0.20}^{+0.17}$ & 1.3 & 1\\
IRDCs & 0.82$_{-0.49}^{+0.76}$ & 6.5 & 2\\
L1544 & 0.74$_{-0.41}^{+0.78}$ & 35 & 3\\
L134N  & 2.2 & 46 & 4\\
TMC-1  & 5.6 & 9 & 4\\
\enddata
\tablenotetext{}{References 1: This study, 2: Miettinen et al. (2011), 3: Redaelli et al. (2019), 4: Tine et al. (2000)}
\end{deluxetable}

\subsubsection{Comparison with Other Regions}

\begin{figure*}
\epsscale{1.0}
\plotone{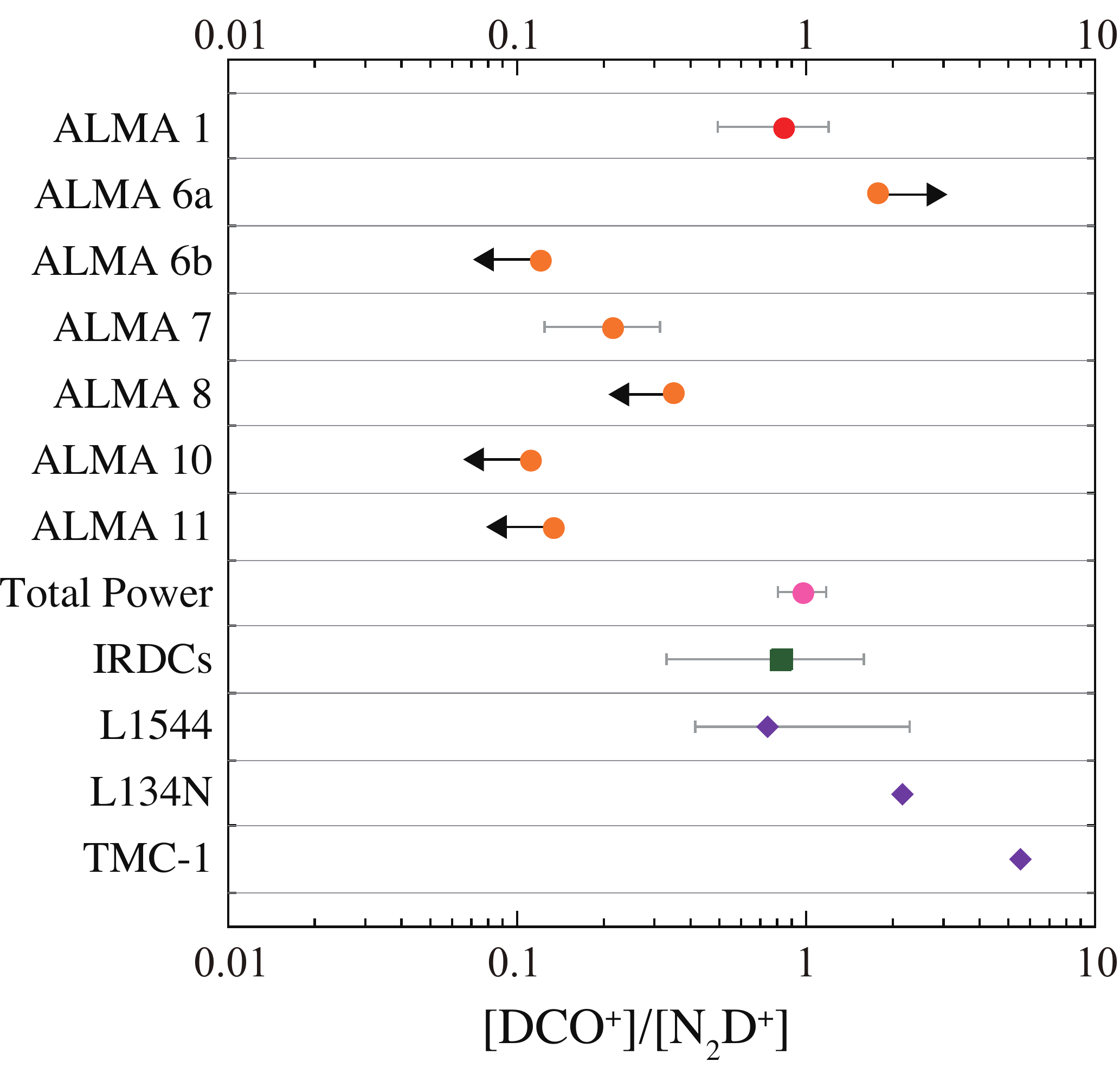}
\caption{DCO$^+$/N$_2$D$^+$ abundance ratio toward the dense cores in G14.49, IRDCs \citep{2011A&A...534A.134M}, L1544 \citep{2019A&A...629A..15R}, L134N \citep{2000A&A...356.1039T}, and TMC-1 \citep{2000A&A...356.1039T}. DCO$^+$/N$_2$D$^+$ abundance ratio derived from the total power data is also plotted. 
Grey bars show the uncertainties, while black arrows show lower and upper limits, respectively. 
\label{fig:dcon2d}}
\end{figure*}

Figure \ref{fig:dcon2d} plots the DCO$^+$/N$_2$D$^+$ abundance ratio of the dense cores in G14.49.
For comparison, we also plot the DCO$^+$/N$_2$D$^+$ abundance ratio of the low-mass starless cores, L1544 \citep{2019A&A...629A..15R}, L134N \citep{2000A&A...356.1039T}, and TMC-1 \citep{2000A&A...356.1039T}. 
The spatial resolution of the observations of the low-mass cores ($\sim$0.01 pc) is comparable to that of our observations ($\sim$0.02 pc).
The DCO$^+$/N$_2$D$^+$ ratio of the dense cores in G14.49, except for ALMA1 and ALMA6a, is found to be lower than those of the low-mass starless cores.
The low DCO$^+$/N$_2$D$^+$ ratio is due to the fact that DCO$^+$ is significantly less abundant than N$_2$D$^+$ in the dense cores in G14.49, as compared with the low-mass starless cores.

Figure \ref{fig:dcon2d} also plots the observation results toward IRDCs taken by \citet{2011A&A...534A.134M}. They observed several IRDCs with APEX, and the value plotted in Figure \ref{fig:dcon2d} is the average value of four IRDCs; the gas kinetic temperature of the four IRDCs ranges from 17 K to 21 K.
The DCO$^+$/N$_2$D$^+$ ratio of the dense cores in G14.49, except for ALMA1 and ALMA6a, appears lower than that of the IRDCs observed by \citet{2011A&A...534A.134M}.

To check whether the difference is due to the different angular resolution between our study and the observations by \citet{2011A&A...534A.134M}, we derive the DCO$^+$/N$_2$D$^+$ ratio from the data of the total power array (the spectra of the total power data are shown in Appendix C). 
Then the DCO$^+$/N$_2$D$^+$ ratio derived from the total power data is found to be comparable to that of the IRDCs observed by \citet{2011A&A...534A.134M} (Table \ref{tbl:DCON2D} and Figure \ref{fig:dcon2d}). Thus, we say that high-angular resolution observations are essential in order to investigate the chemical compositions of the dense cores, because the IRDC clumps contain the dense cores with various evolutionary stages.

\subsubsection{Comparison with Simple Chemical Model Calculations}

\begin{figure*}
\epsscale{1.2}
\plotone{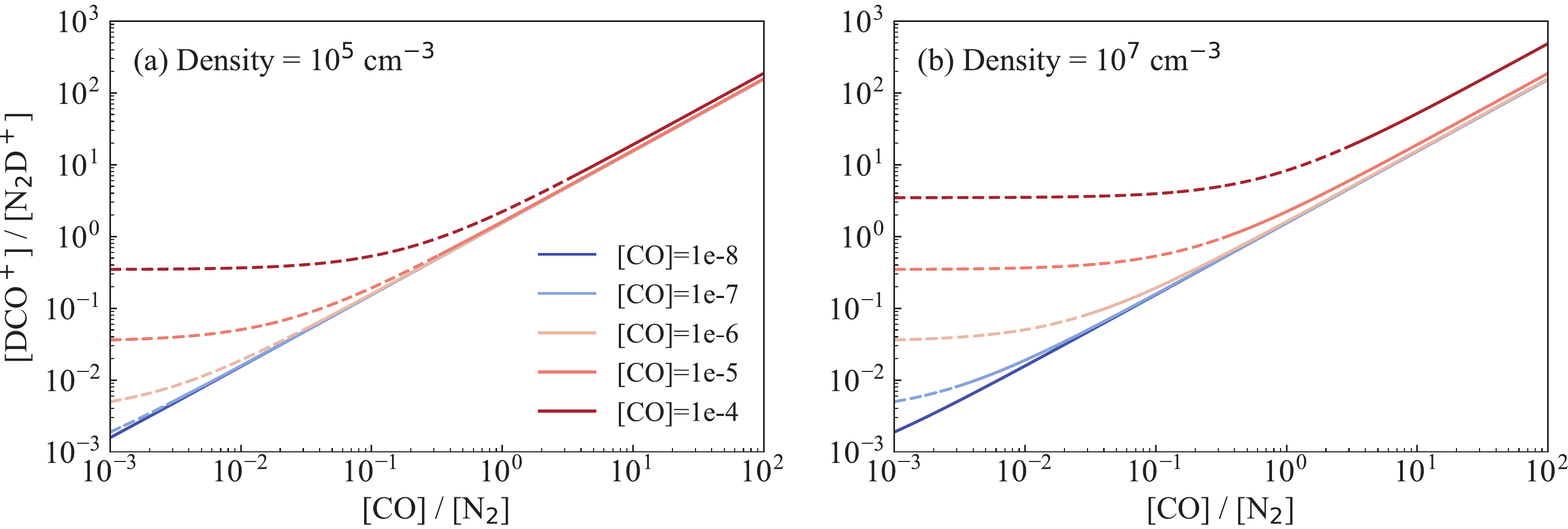}
\caption{DCO$^+$/N$_2$D$^+$ abundance ratio as a function of the CO/N$_2$ abundance ratio given by Eq. \ref{eq:dco+_n2d+}. 
The left and right panels represent the cases when $n_{\rm H}$ = 10$^5$ cm$^{-3}$ and $n_{\rm H}$ = 10$^7$ cm$^{-3}$, respectively.
Temperature is set to be 20 K.
As the cosmic abundance of elemental nitrogen is $6\times10^{-5}$ \citep{Przybilla08}, the maximum abundance of N$_2$ is $3\times10^{-5}$.
Then, for example, the case when the CO abundance is 10$^{-4}$ and the CO/N$_2$ abundance ratio is unity is unrealistic.
Such regions are shown by dashed lines.
\label{fig:analytical}
}
\end{figure*}

As mentioned in the previous section, DCO$^+$ and N$_2$D$^+$ are formed from CO and N$_2$, respectively. Thus, the abundances of DCO$^+$ and N$_2$D$^+$ are thought to depend on the abundance of CO and N$_2$. 
If so, a low DCO$^+$/N$_2$D$^+$ ratio could reflect a low CO/N$_2$ ratio in the dense cores.
Considering the balance between formation and destruction of N$_2$D$^+$, the N$_2$D$^+$ abundance relative to hydrogen nuclei ([N$_2$D$^+$]) is given by
\begin{equation}
[{\rm N_2D^+}] = k_{\ref{react:n2d+_p}}[{\rm N_2}][{\rm H_2D^+}]/(k_{\ref{react:n2d+_d1}}[{\rm CO}] + k_{\ref{react:n2d+_d2}}[{\rm e^-}]),
\end{equation}
where [e$^-$] is the electron abundance relative to hydrogen nuclei, and $k_{\ref{react:n2d+_p}}$, $k_{\ref{react:n2d+_d1}}$, and $k_{\ref{react:n2d+_d2}}$ are 
the rate coefficients of Reactions \ref{react:n2d+_p}, \ref{react:n2d+_d1}, and \ref{react:n2d+_d2}, respectively.
Similarly, the DCO$^+$ abundance is given by
\begin{equation}
[{\rm DCO^+}] = (k_{\ref{react:dco+_p}}[{\rm CO}][{\rm H_2D^+}] + k_{\ref{react:n2d+_d1}}[{\rm CO}][{\rm N_2D^+}])/k_{\ref{react:dco+_d}}[{\rm e^-}].
\end{equation}
Then the DCO$^+$/N$_2$D$^+$ abundance ratio is given by
\begin{equation}
%\frac{[{\rm DCO^+}]}{[{\rm N_2D^+}]} = \beta \left( k_2k_4 \frac{[{\rm CO}]^2}{[{\rm e^-}][{\rm N_2}]} +  k_3 k_4 \frac{[{\rm CO}]}{[{\rm N_2}]} +  k_1k_2\frac{[{\rm CO}]}{[{\rm e^-}]}\right),
\frac{[{\rm DCO^+}]}{[{\rm N_2D^+}]} = \beta \left( \frac{k_{\ref{react:n2d+_d1}}}{k_{\ref{react:n2d+_d2}}} \frac{[{\rm CO}]}{[{\rm e^-}]} +  1 \right) \frac{[{\rm CO}]}{[{\rm N_2}]}  +  
\frac{k_{\ref{react:n2d+_d1}}}{k_{\ref{react:dco+_d}}}\frac{[{\rm CO}]}{[{\rm e^-}]}, \label{eq:dco+_n2d+}
\end{equation}
where $\beta$ is given by $k_{\ref{react:n2d+_d2}}k_{\ref{react:dco+_p}}/(k_{\ref{react:n2d+_p}} k_{\ref{react:dco+_d}})$.
Figure \ref{fig:analytical} visualizes Eq. \ref{eq:dco+_n2d+} as functions of [CO]/[N$_2$], varying [CO].
In the figure, [e$^-$] is evaluated by $\sqrt{2\xi/(\alpha n_{\rm H})}$ \citep[e.g.,][]{tielens05}, 
where $\xi$ is the cosmic-ray ionization rate ($2.6\times10^{-17}$ s$^{-1}$), $\alpha$ is the electron recombination rate of H$_3^+$, 
and $n_{\rm H}$ is the volume density of hydrogen nuclei; 
this should be valid in regions where UV radiation field is attenuated and the cosmic-rays are the dominating ionization source.
The value used for [e$^-$] is 1.6$\times$10$^{-7}$ and 1.6$\times$10$^{-8}$ for 10$^5$ cm$^{-3}$ and 10$^7$ cm$^{-3}$, respectively.
All the rate coefficients were taken from the KIDA (KInetic Database for Astrochemistry) database \citep{wakelam15} \footnote{http://kida.obs.u-bordeaux1.fr}.
%When $[{\rm N_2}]/[{\rm e^-}] < k_3k_4/(k_1k_2)$, i.e., $1000[{\rm e^-}] > [{\rm N_2}]$, the second term dominates over the third term.
%When $[{\rm CO}]/[{\rm e^-}] < k_3/k_2$, i.e., $1000[{\rm e^-}] < [{\rm CO}]$, the second term dominates over the first term.
It is clear that in order to explain the low DCO$^+$/N$_2$D$^+$ ratio ($\lesssim$0.1) in the dense cores in category 2, 
both a low CO abundance ($\lesssim$10$^{-6}$) and low [CO]/[N$_2$] ratio  ($\lesssim$0.1) are required (see Figure \ref{fig:analytical}). 
Note that the plots in Figure \ref{fig:analytical} change very little within a temperature range of 10-30 K, because the model considers only the gas-phase reactions.

Since the sublimation temperature of CO ($\sim$25 K) is known to be higher than that of N$_2$ ($\sim$20 K) \citep{2016ApJ...816L..28F}, the CO/N$_2$ ratio is considered to depend on temperature.
In order to achieve a low CO/N$_2$ ratio, the temperature of the dense core would need to be higher than the sublimation temperature of N$_2$ and lower than the sublimation temperature of CO. In that case, N$_2$ would sublimate from the surface of the grains, while CO would stay on them in ice form, and consequently, the CO/N$_2$ ratio would be low in the gas phase.

\subsubsection{Chemical Network Calculations}

To verify the above simple scenario, we perform more detailed chemical model calculations.
The chemical reaction network used is the same as the one used in \citet{2015ApJ...803...70S}. 
Briefly, the chemical model adopts a three-phase model, in which gas, an ice surface, and an ice mantle are considered \citep{1993MNRAS.263..589H, 2015A&A...584A.124F}. We treat the top four monolayers of the ice as the surface, following \citet{2013ApJ...762...86V}, while the rest as the ice mantle. As chemical processes, the model includes gas-phase reactions, interactions between gas-phase and the surface, surface reactions, and reactions in ice mantles. Swapping between the surface active layer and the inner ice mantle (i.e., the thermal diffusion of ice mantle) is also considered following the prescription proposed by \citet{2013ApJ...765...60G}. We set the activation energy barrier for swapping to be 1.2$E_{\rm des}$, where $E_{\rm des}$ is the adsorption energy. The cosmic-ray ionization rate is set to be 2.6$\times$10$^{-17}$ s$^{-1}$ \citep{2000A&A...358L..79V}).
As non-thermal desorption processes, our model includes photodesorption \citep[e.g.,][]{2011ApJ...739L..36F, 2013A&A...556A.122F}, chemical desorption \citep{2007A&A...467.1103G}, and stochastic heating by cosmic-rays \citep{1993MNRAS.261...83H}, the last of which is the most important for the non-thermal desorption of CO and N$_2$ in our simulations.

\begin{figure*}
\epsscale{0.7}
\plotone{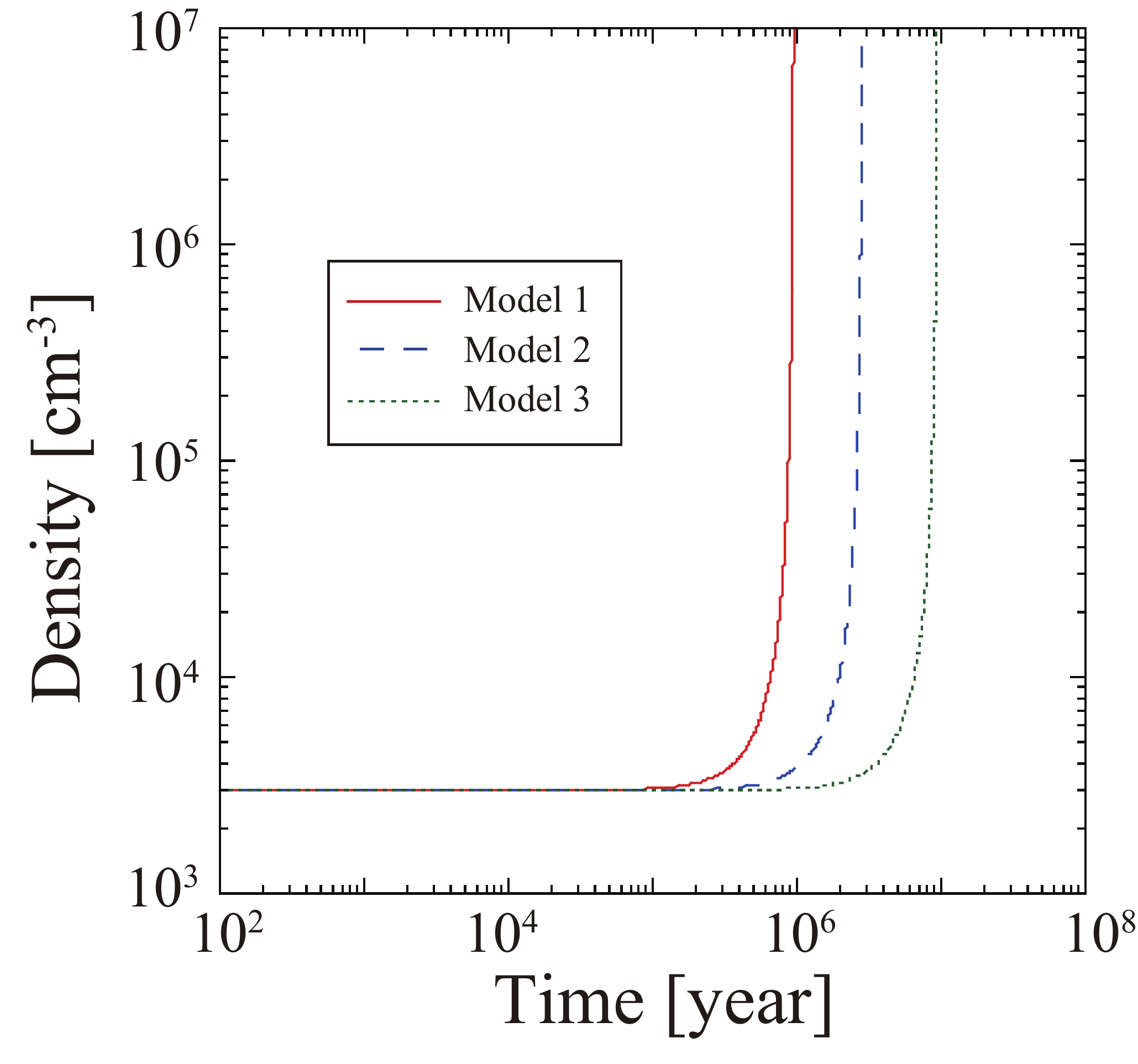}
\caption{Time dependence of the density used for Model 1 (red line), Model 2 (blue dashed line), and Model 3 (green dotted line).\label{fig:chmodeltvsn}}
\end{figure*}

At first, we calculate the chemical composition of the collapsing starless core phase, where the volume density increases over time, using Equation (2) in \citet{1988MNRAS.231..409B}. We calculate the chemical composition of dense cores with three different collapsing speeds; $Model$ 1: free fall time, $Model$ 2: three times slower than free-fall time, and $Model$ 3: ten times slower than free fall time.
In all models, the initial and final volume densities are set to 2$\times$10$^{3}$ cm$^{-3}$ and 10$^{7}$ cm$^{-3}$, respectively.
Figure \ref{fig:chmodeltvsn} shows the time dependence of the density for models 1,2, and 3.
In the calculations, we focus on a fluid fragment of a collapsing core, and solve for the chemical evolution of the fragment.
Thus, the geometric structure of the core is not taken into account.
As an initial condition, all the elements are assumed to be in the form of neutral atom or atomic ion, depending on their ionization potentials, except for hydrogen and deuterium, which are in molecular form (H$_2$ and HD).
In the collapsing starless phase, gas and dust temperatures are fixed to be 10 K.

\begin{figure*}
\epsscale{1.2}
\plotone{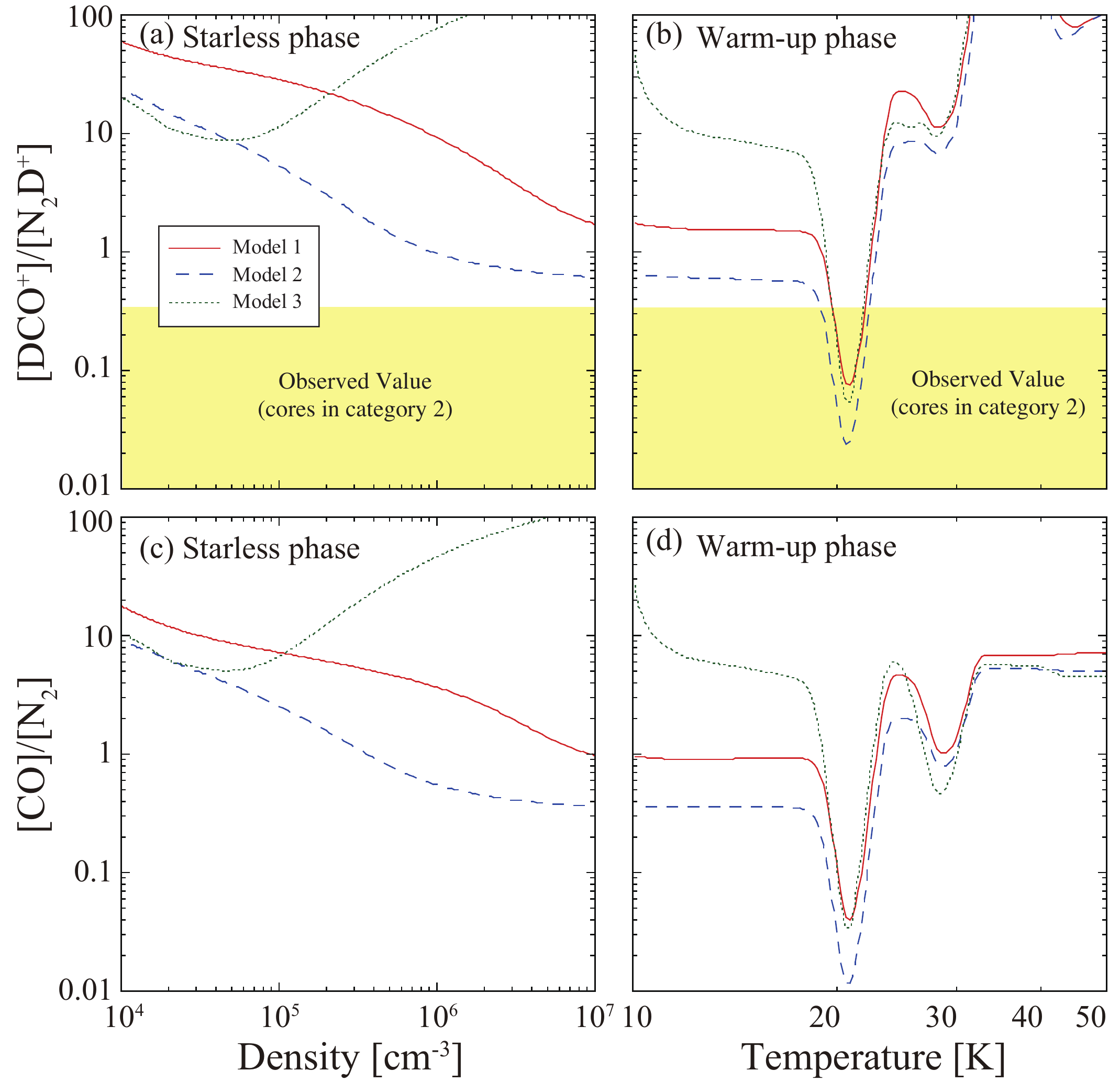}
\caption{(a) Chemical model calculation results of DCO$^+$/N$_2$D$^+$ ratio in starless phase. The horizontal axis is the density, which increases with time, and the vertical axis is the DCO$^+$/N$_2$D$^+$ ratio. The red line indicates the result of Model 1 (collapsing with free fall time), and the blue dashed line indicates the result of Model 2 (collapsing three times slower than the free-fall time), and the green dotted line indicates the result of Model 3 (collapsing ten times slower than the free-fall time). (b) Chemical model calculation results of DCO$^+$/N$_2$D$^+$ ratio in the warm-up phase. The horizontal axis is temperature, which increases with time, and the vertical axis is the DCO$^+$/N$_2$D$^+$ ratio. (c) same as (a), but for CO/N$_2$ ratio. (d) same as (b), but for CO/N$_2$ ratio.\label{fig:chmodel}}
\end{figure*}

Figure \ref{fig:chmodel}a and Figure \ref{fig:chmodel}c show the model calculation results of the DCO$^+$/N$_2$D$^+$ and CO/N$_2$ abundance ratios in the starless phase, respectively.
The behavior of the CO/N$_2$ ratio is found to be very similar to that of the DCO$^+$/N$_2$D$^+$ ratio.
Thus, the DCO$^+$/N$_2$D$^+$ ratio of starless cores is most likely to reflect the CO/N$_2$ ratio.

\begin{figure*}
\epsscale{0.9}
\plotone{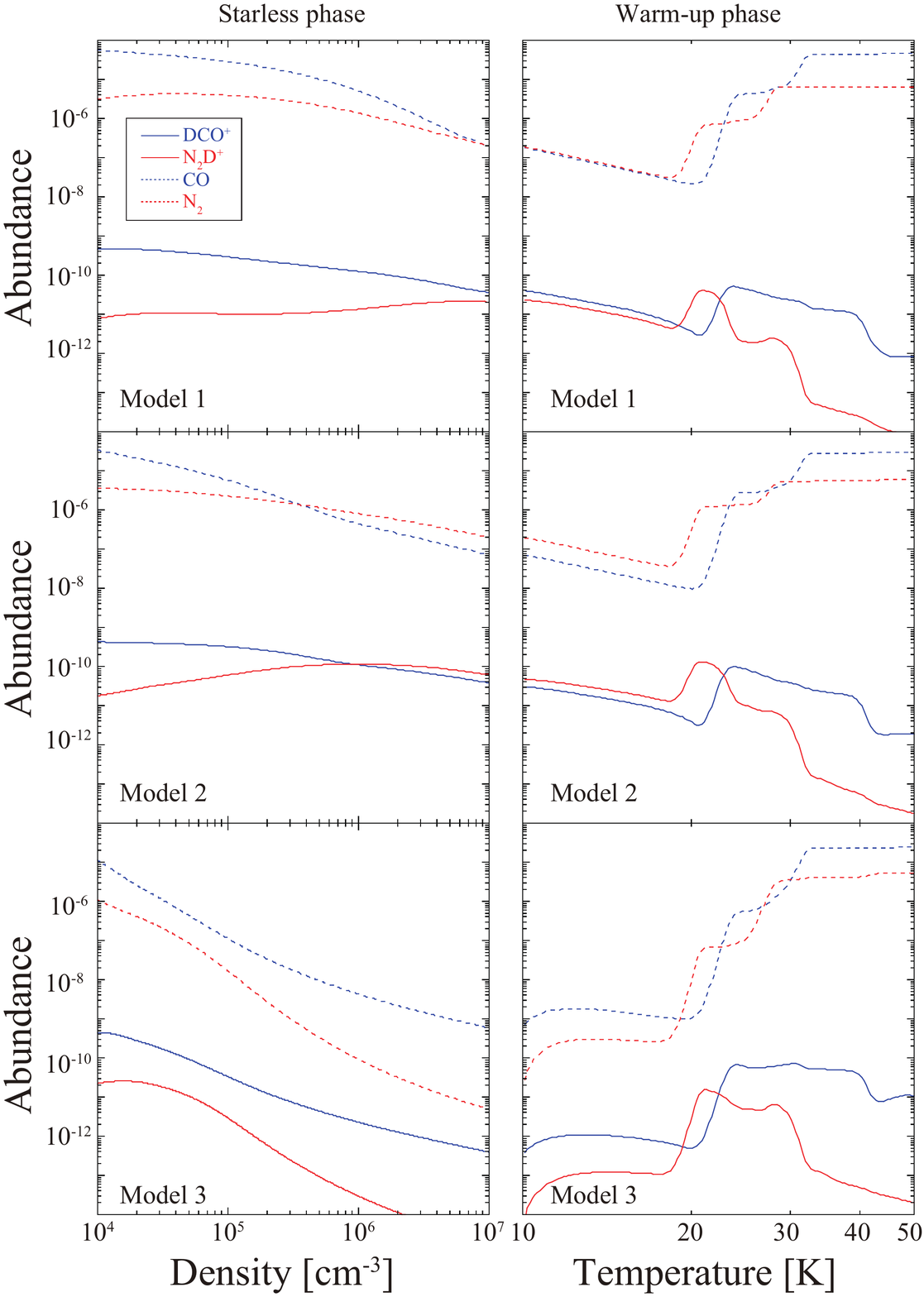}
\caption{Chemical model calculation results of the abundances of N$_2$D$^+$, DCO$^+$, CO and N$_2$. Left panels show the results for the starless phase. The horizontal axis is the density, which increases with time. Right panels show the results for the warm-up phase. The horizontal axis is temperature, which increases with time. Top: Model 1 (collapsing with free fall time), Middel: Model 2 (collapsing three times slower than the free-fall time), and bottom: Model 3 (collapsing ten times slower than the free-fall time). \label{fig:chmodel2}}
\end{figure*}

In Figure \ref{fig:chmodel}a, we can see that the DCO$^+$/N$_2$D$^+$ ratio at a given density depends on the collapsing speed.
In Models 1 and 2, both the DCO$^+$/N$_2$D$^+$ and the CO/N$_2$ ratio decreases with increasing densities. 
This is due to the effect of the non-thermal desorption. Although CO and N$_2$ freeze out on grain surfaces in the dense regions, some of them can go back to the gas phase due to the non-thermal desorption. In addition, since the binding energy of N$_2$ is lower than that of CO, N$_2$ is desorbed more efficiently than CO by stochastic heating.
As a result, the gas-phase CO/N$_2$ ratio decreases in the dense regions.
Thus, the higher DCO$^+$/N$_2$D$^+$ ratio of low-mass starless cores could be due to lower volume densities than those observed in the dense cores of G14.49.
Although the DCO$^+$/N$_2$D$^+$ ratio increases with increasing the density in Model 3, most of the molecules are heavily depleted onto grain surfaces (see the left bottom panel of Figure \ref{fig:chmodel2}). As seen in Figure \ref{fig:chmodeltvsn}, the core in Model 3 experiences high-density regions for a longer time than that in Models 1 and 2, so that many molecules can freeze onto grain surfaces significantly in Model 3. Since the DCO$^+$ emission is strong toward the low-mass starless cores, Model 3 is not the case for low-mass starless cores.

In Figure \ref{fig:chmodel}a, it also appears that all calculated DCO$^+$/N$_2$D$^+$ ratios are higher than the observed values for the cores in category 2, except for ALMA6a. Thus, the observed DCO$^+$/N$_2$D$^+$ ratios cannot be reproduced by chemical model calculations with a constant temperature of 10 K.

Next, we simulate the chemical composition after the onset of star formation (i.e., warm-up phase) in order to investigate whether the low DCO$^+$/N$_2$D$^+$ ratio really appears in the temperature range from 20 to 25 K. 
The temperature increases from 10 to 200 K within a timescale of 10$^4$ yr. A constant volume density of 10$^7$ cm$^{-3}$ is assumed, because the density derived from the dust continuum is typically 10$^7$ cm$^{-3}$ (Table \ref{tbl:CoreP}). 
The model calculation results of the starless phase at a density of 10$^7$ cm$^{-3}$ are set as the initial abundances for the warm-up phase.

Figure \ref{fig:chmodel}b shows the result of the DCO$^+$/N$_2$D$^+$ ratio in the warm-up phase.
In Figure \ref{fig:chmodel}b, we can see a dip of DCO$^+$/N$_2$D$^+$ ratio in a temperature range of 20-25 K.
The DCO$^+$/N$_2$D$^+$ ratio in this temperature range is comparable to that of the observed values for the cores in category 2, except for ALMA6a.
In Figures \ref{fig:chmodel}d, we plot the model calculation results of CO/N$_2$ for the warm-up phase.
The CO/N$_2$ ratio behaves very similar to the DCO$^+$/N$_2$D$^+$ ratio in temperatures of less than 25 K, and we can see a dip of CO/N$_2$ ratio in the temperature range of 20-25 K. In this temperature range, CO is depleted onto grain surface, while N$_2$ sublimates from grain surfaces, as seen in the right panels of Figure \ref{fig:chmodel2}.
Thus, the low DCO$^+$/N$_2$D$^+$ ratio observed toward the dense cores in G14.49 is likely to reflect the low CO/N$_2$ ratio of the dense cores. 

In Figures \ref{fig:chmodel}b and \ref{fig:chmodel}d, we can also see a slight dip at a temperature of 30 K. This dip is due to the difference in the energy barrier of the swapping from the inner ice mantle to the surface layer between CO and N$_2$.
The energy barrier of the swapping is assumed to be higher than the barrier of the thermal desorption in the chemical model.
Thus, CO and N$_2$ can remain in the inner ice mantle even above their sublimation temperatures.
With increasing temperatures, CO and N$_2$ in the inner ice mantles come to the surface layer and are thermally desorbed. As a result, the gas-phase abundance of CO and N$_2$ can increase two steps, as seen in Figure \ref{fig:chmodel2}. The second dip is due to the difference in the temperature of the second step.

In Figure \ref{fig:chmodel}b, the DCO$^+$/N$_2$D$^+$ ratio increases with increasing the temperature above the temperature of 30 K.
This is because N$_2$D$^+$ is destroyed efficiently in warm regions due to the reaction with CO or electron.

In summary, the observed low DCO$^+$/N$_2$D$^+$ ratio toward the dense cores in G14.49 could be due to higher temperature and density of the dense cores in G14.49 than those of the low-mass starless cores.
Although the temperature derived from the infrared data is about 13 K \citep{2019ApJ...886..102S}, the single-dish observations used for the SED fitting (Herschel/HiGAL and APEX/ATLASGAL) likely trace the low-density envelope due to low angular resolution (35$^{\prime\prime}$ at 500 $\mu$m).
Since there are many protostars in this clump, the temperature of dense regions is likely higher than that of the envelope. 
For a more accurate comparison with the chemical models, it is important to reveal the temperature of the emitting regions by, for example, multi-line observations or multi-wavelength observations of the continuum emission.

\subsection{Implications}

According to the simple consideration of the difference of the desorption temperature between CO and N$_2$ as well as the detailed chemical model calculation, the low DCO$^+$/N$_2$D$^+$ ratio likely originates from the lukewarm temperature condition (20-25 K). In the chemical model calculations, we assume the temporal raise of the temperature due to protostellar heating. If this is the case, the DCO$^+$/N$_2$D$^+$ ratio would reflect the evolutionary stage of the dense cores. However, the heating source can also be external to produce the lukewarm condition. 
In fact, there is no evidence of an embedded protostar toward ALMA11. 
In addition, the N$_2$D$^+$ emission is rather widely spread out in the dense cores, while the DCO$^+$ emission is not.
In order to heat large regions of the core, the heating source should be luminous, and the inner part should be hotter than the envelope. However, we cannot find any evidence of a protostar in this core.
Therefore, the sources of heating would not be internal, but should be external.

A molecular outflow, emanating from ALMA1, is in the immediate vicinity of ALMA11 \citep{Li20b}. Thus, the temperature of ALMA11 could be externally enhanced by shocks from the molecular outflow. Since there are two outflows, emanating from ALMA2 and ALMA29, near ALMA7, ALMA7 may also be heated by the outflows.

If the low DCO$^+$/N$_2$D$^+$ ratio is achieved in the narrow temperature range (20-25 K), this result raises an important question of whether the dense cores with a low DCO$^+$/N$_2$D$^+$ ratio is also found in other IRDCs.
To investigate this issue, statistical studies of DCO$^+$/N$_2$D$^+$ ratio toward 70 $\mu$m-dark IRDCs are interesting.
In addition, we note that observations of cold gas tracers, such as H$_2$D$^+$, are also crucial to understanding the properties of cold ($<$10 K) dense cores where molecules are heavily depleted onto grain surfaces.

\section{Summary}

We summarize the results below.

\begin{itemize} 

\item The distribution is found to be different among N$_2$D$^+$, DCO$^+$, and DCN.
We suggest that it is caused by the difference of the sublimation temperature of N$_2$, CO, and DCN (HCN) as well as the formation and destruction processes. 

\item The C$^{18}$O emission does not peak toward most of the dust continuum peaks. It does not have a peak toward which the N$_2$D$^+$ emission is strong, either. It indicates that CO is depleted onto dust grains.

\item The velocity width of C$^{18}$O is found to be much broader than those of the deuterated molecular lines, and the velocity width of the deuterated molecular lines is as narrow as that of the cores in low-mass star-forming regions.
This could suggest that a turbulent low-density envelope encloses the less turbulent dense cores.

\item We found that the DCO$^+$/N$_2$D$^+$ abundance ratio of the dense cores in category 2 is lower than the ratios of starless cores from low-mass star-forming regions. In the G14.49 cores, DCO$^+$ is significantly less abundant than N$_2$D$^+$, which is remarkable once compared with low-mass sources.
We carried out chemical model calculations and found that the DCO$^+$/N$_2$D$^+$ ratio depends on the CO/N$_2$ ratio, and the low DCO$^+$/N$_2$D$^+$ ratios can be reproduced in the temperature range between the sublimation temperature of N$_2$ and CO, which is 20-25 K. Thus, the low DCO$^+$/N$_2$ ratio suggests that the cores in G14.49 are warmer and denser than the cores in the low-mass star-forming regions. 

\end{itemize} 

\acknowledgments
We would like to thank the anonymous referee for her/his useful comments to improve our manuscript.
P.S. was partially supported by a Grant-in-Aid for Scientific Research (KAKENHI Number 18H01259) of the Japan Society for the Promotion of Science (JSPS). Data analysis was in part carried out on the Multi-wavelength Data Analysis System operated by the Astronomy Data Center (ADC), National Astronomical Observatory of Japan. A.G. gratefully acknowledges the support from the NAOJ Visiting Fellow Program to visit the National Astronomical Observatory of Japan in November-December 2016. 
Y.A. acknowledges support by NAOJ ALMA Scientific Research Grant Numbers 2019-13B.
This paper makes use of the following ALMA data: ADS/JAO.ALMA\#2015.1.01539.S. ALMA is a partnership of ESO (representing its member states), NSF (USA) and NINS (Japan), together with NRC (Canada), MOST and ASIAA (Taiwan), and KASI (Republic of Korea), in cooperation with the Republic of Chile. The Joint ALMA Observatory is operated by ESO, AUI/NRAO and NAOJ.
This study is supported by KAKENHI (20K04025, 20H05645, 20H05845, 20H5847, and 18H05222).


\vspace{5mm}
\facilities{ALMA}


\software{CASA \citep{2007ASPC..376..127M}}
\software{yclean \citep{2018ApJ...861...14C}}



\clearpage

\appendix

\restartappendixnumbering

\section{Parameters of Dense Cores in G14.49}\label{sec:dcol}

The parameters of the eleven dense cores with the high H$_2$ column density are listed in Table \ref{tbl:CoreP}.
Details about the parameters are described in \citet{2019ApJ...886..102S}.
We note that the positions of the cores listed in Table \ref{tbl:CoreP} are the peak positions of the continuum emission, which are different from the positions listed in \citet{2019ApJ...886..102S}.
They listed the positions of the centroid derived by the dendrogram method.

\begin{deluxetable}{crrrrrr}
\tabletypesize{\scriptsize}
\tablecaption{Parameters of the cores in G14.49\label{tbl:CoreP}}
\tablewidth{0pt}
\tablehead{
%\colhead{1} & \colhead{1} & \colhead{1} & \colhead{1} & \colhead{1} & \colhead{1} &  \\
\colhead{Core ID} & \colhead{R.A.(J2000)} & \colhead{Dec.(J2000)} & \colhead{$N_{\rm peak}$(H$_2$)} & \colhead{Mass} & \colhead{Density} & \colhead{Category\tablenotemark{a}} \\
 & & & \footnotesize[10$^{23}$ cm$^{-2}$] & \footnotesize[$M_\odot$] & \footnotesize[10$^{6}$ cm$^{-3}$] & \\
}
\startdata
%1 & 1 & 1 & 1 & 1 & 1 & 1 \\
ALMA1  & 18$^{\rm h}$17$^{\rm m}$22.56$^{\rm s}$ & -16$^{\circ}$25$^{\prime}$00.7$^{\prime\prime}$ & 6.29 & 20.71 & 13.5 & 3\\
ALMA2  & 18$^{\rm h}$17$^{\rm m}$21.43$^{\rm s}$ & -16$^{\circ}$25$^{\prime}$02.7$^{\prime\prime}$ & 5.66 & 10.43 & 20.8 & 3 \\
ALMA3  & 18$^{\rm h}$17$^{\rm m}$22.11$^{\rm s}$ & -16$^{\circ}$25$^{\prime}$05.3$^{\prime\prime}$ & 4.00 & 9.54 & 7.7 & 3\\
ALMA4  & 18$^{\rm h}$17$^{\rm m}$22.86$^{\rm s}$ & -16$^{\circ}$24$^{\prime}$55.1$^{\prime\prime}$ & 3.41 & 5.44 & 16.1 & 3\\
ALMA5  & 18$^{\rm h}$17$^{\rm m}$21.54$^{\rm s}$ & -16$^{\circ}$24$^{\prime}$58.7$^{\prime\prime}$ & 3.04 & 3.52 & 18.5 & 2 \\
ALMA6  & 18$^{\rm h}$17$^{\rm m}$21.65$^{\rm s}$ & -16$^{\circ}$25$^{\prime}$03.7$^{\prime\prime}$ & 2.52 & 2.07 & 26.2 & 2\\
ALMA7  & 18$^{\rm h}$17$^{\rm m}$21.32$^{\rm s}$ & -16$^{\circ}$24$^{\prime}$56.1$^{\prime\prime}$ & 2.24 & 5.99 & 9.4 & 2 \\
ALMA8  & 18$^{\rm h}$17$^{\rm m}$22.50$^{\rm s}$ & -16$^{\circ}$24$^{\prime}$58.1$^{\prime\prime}$ & 2.17 & 2.47 & 17.8 & 2\\
ALMA9  & 18$^{\rm h}$17$^{\rm m}$21.46$^{\rm s}$ & -16$^{\circ}$24$^{\prime}$54.5$^{\prime\prime}$ & 2.14 & 2.17 & 14.2 & 0 \\
ALMA10 & 18$^{\rm h}$17$^{\rm m}$21.39$^{\rm s}$ & -16$^{\circ}$24$^{\prime}$59.3$^{\prime\prime}$ & 2.10 & 4.48 & 9.8 & 2\\
ALMA11 & 18$^{\rm h}$17$^{\rm m}$22.39$^{\rm s}$ & -16$^{\circ}$24$^{\prime}$56.7$^{\prime\prime}$ & 1.93 & 4.05 & 10.3 & 2\\
\enddata
\tablenotetext{a}{Categories are classified by \citet{2019ApJ...886..102S}.}
\end{deluxetable}

\section{Derivation of Column density}\label{sec:dcol}

Assuming the optically thin condition and a beam filling factor of 1, we derive the column density by using the following equation;
\begin{equation}
N_{total}=\frac{3h}{8\pi^3 \mu^2 S }Q \frac{\exp\left(\frac{E_u}{kT_{\rm ex}}\right)}{\exp\left(\frac{h \nu}{kT_{\rm ex}}\right)-1} \frac{W}{J(T_{\rm ex})-J(T_{\rm BB})}, \label{eq:col}
\end{equation}
where $h$ is the Planck constant, $Q$ is the partition function, $\nu$ is the rest frequency, $\mu$ is the electric dipole moment, $S$ is the line strength, $E_u$ is the upper state energy, $T_{\rm ex}$ is the excitation temperature, $T_{\rm BB}$ is the temperature of the cosmic background radiation (2.73 K), $W$ is the integrated intensity, $k$ is the Boltzmann constant, and $J$ stands for the Planck function for the given temperature and frequency ($\nu$) as:
\begin{equation}
J(T) = \frac{\frac{h\nu}{k}}{\exp\left(\frac{h\nu}{kT}\right)-1}.
\end{equation}

The partition function $Q$ is evaluated from the data in the Cologne Database for Molecular Spectroscopy \citep[CDMS;][]{2005JMoSt.742..215M}. We fit the CDMS data to the function of $Q$($T$) = $a$$\times$$T^b$, where $a$ and $b$ are fitting parameters. For N$_2$D$^+$, ($a$, $b$)=(4.957, 0.99724). For DCO$^+$, ($a$, $b$)=(0.59878, 0.99428).

The integrated intensity is calculated from peak intensity and velocity width derived by the gaussian fitting.
We use the following relation between the main beam temperature ($T_{MB}$) and flux ($S_\nu$),
\begin{equation}
T_{MB} = \frac{c^2}{2k\nu^2 \Omega_B}S_{\nu},
\end{equation} 
where $\nu$ is the rest frequency, $\Omega_{\rm B}$ is a solid angle of the beam. 
Then, the integrated intensity is derived by the following equation,
\begin{equation}
W = \frac{c^2}{2k\nu^2 \Omega_B} \left( \frac{1}{2} \sqrt{\frac{\pi}{\ln{2}}} \right) S_{pk} \Delta V,
\end{equation}
where $c$ is the speed of light, $S_{pk}$ is the peak flux density, $\Delta V$ is the velocity width, and the factor of $\sqrt{\pi/\ln{2}}/2$ is for the integration over a Gaussian profile.

\section{Spectra taken with the total power array}\label{sec:dcol}

\begin{figure}
\epsscale{0.3}
\plotone{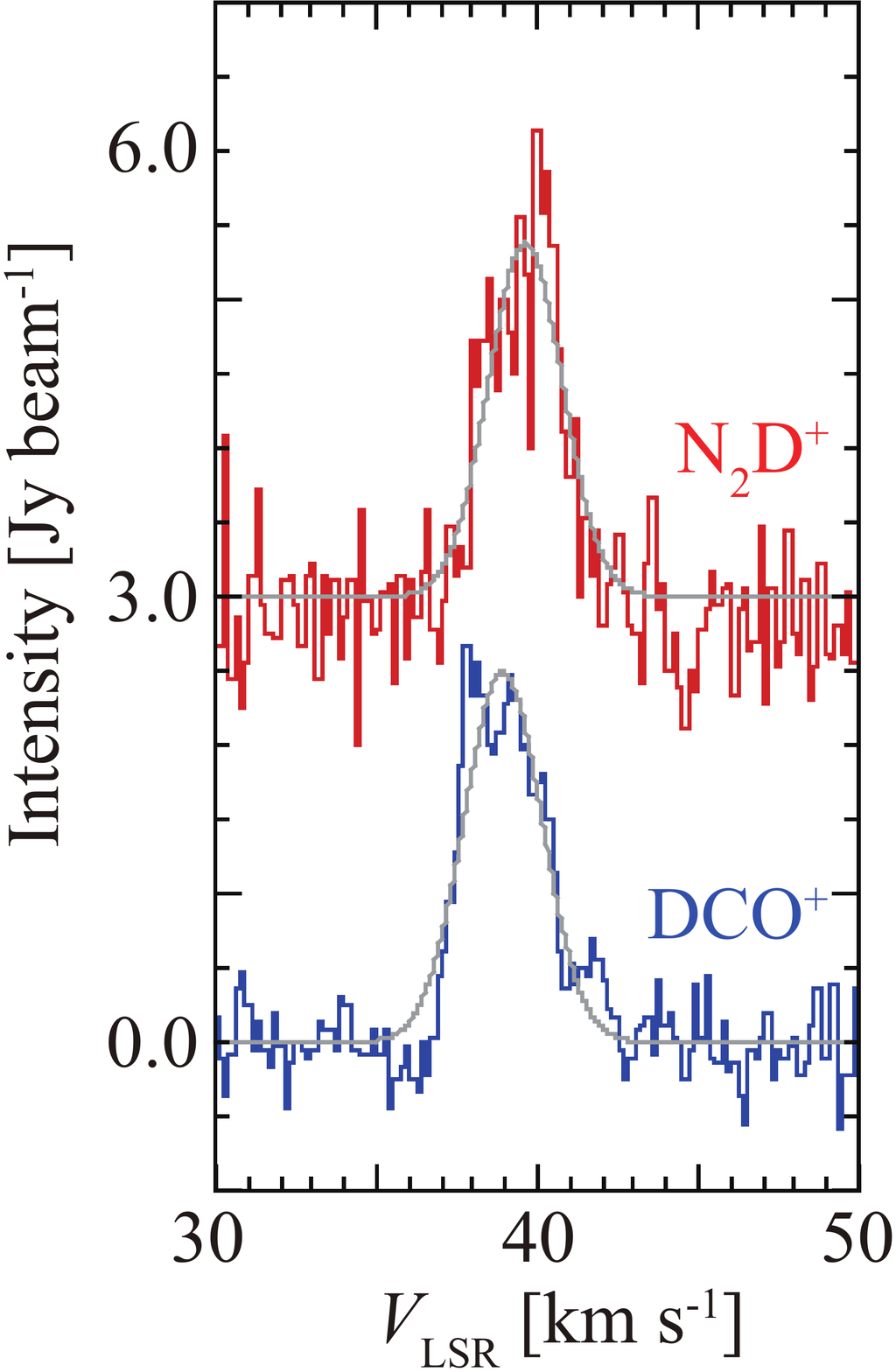}
\caption{Spectra of N$_2$D$^+$ and DCO$^+$ taken with the total power array toward a position of (R.A.(J2000), Dec.(J2000)) = (18$^{\rm h}$17$^{\rm m}$22.0$^{\rm s}$, -16$^{\circ}$25$^{\prime}$1.9$^{\prime\prime}$). \label{fig:spectp}}
\end{figure}

Figure \ref{fig:spectp} shows the spectra of N$_2$D$^+$ and DCO$^+$ taken with the total power array toward the center of this clump, (R.A.(J2000), Dec.(J2000)) = (18$^{\rm h}$17$^{\rm m}$22.0$^{\rm s}$, -16$^{\circ}$25$^{\prime}$1$^{\prime\prime}$.9). The beam size of the total power data is 28$^{\prime\prime}$.1 and 30$^{\prime\prime}$.1 for DCO$^+$ and N$_2$D$^+$, respectively.
The results of the single Gaussian fitting are listed in Table \ref{tbl:SpTp}. 

\clearpage

\begin{deluxetable}{crrr}
\tabletypesize{\scriptsize}
\tablecaption{Line Parameters of N$_2$D$^+$ and DCO$^+$ of the total power array data\label{tbl:SpTp}}
\tablewidth{0pt}
\tablehead{
 &\colhead{$S_{pk}$}  & \colhead{$\Delta V$} & \colhead{$V_{\rm LSR}$} \\
 & \footnotesize[Jy beam$^{-1}$] & \footnotesize[km s$^{-1}$]& \footnotesize[km s$^{-1}$]\\
}
\startdata
N$_2$D$^+$ &2.38${\pm 0.14}$ &  2.72${\pm 0.18}$ & 39.65${\pm 0.08}$  \\
DCO$^+$ &  2.49${\pm 0.08}$ &  2.85${\pm 0.11}$ & 38.95${\pm 0.05}$  \\
\enddata
\tablenotetext{}{$S_{pk}$: peak flux density; $\Delta V$: full width per half maximum}
\end{deluxetable}


%\bibliography{}{}
%\bibliographystyle{aasjournal}



%\begin{figure}
%\plotone{Figure6.eps}
%\caption{Chemical model calculations.\label{fig:fig4}}
%\end{figure}

\clearpage

\end{document}